\documentclass[11pt]{article} 

\usepackage[utf8]{inputenc} 
\usepackage{geometry} 

\usepackage{graphicx} 
\usepackage{lineno}

\usepackage{amsmath}
\usepackage{amsfonts}
\usepackage[numbers]{natbib}
\usepackage{booktabs} 
\usepackage{array} 
\usepackage{paralist} 
\usepackage{verbatim} 
\usepackage{subfig} 

\usepackage{chemist}
\usepackage{multicol}
\usepackage{multirow}
\usepackage[title]{appendix}

\usepackage{fancyhdr} 
\usepackage[font={small,it}]{caption}
\usepackage{sectsty}
\allsectionsfont{\sffamily\mdseries\upshape} 

\usepackage[nottoc,notlof,notlot]{tocbibind} 
\usepackage[titles,subfigure]{tocloft} 

\title{Intermolecular Enzymatic  Encoding of Nucleic Acid, Steroid Complexes: A New Theory on the Chemical Origin of Life Based on Evidence of Structural Symmetry}
\author{Charles D. Schaper, Ph.D.\\cschaper@transferdevices.com\\Union City, California, USA}
\date{July 5, 2020}

\begin{document}
\maketitle

\begin{abstract}
The origin of life is one of the greatest mysteries. The mechanism for the synthesis of DNA is synonymous with the chemical origin of life, and theories have been developed along many lines of reasoning, but resolving all requirements remains a challenge, such as defining an objective path to produce sequences of encoded nucleotides paired as adenine to thymine and guanine to cytosine. Here, a new theory for the origin of DNA is presented. The theory is based upon three lines of experimental evidence and agreement of structural symmetry between DNA nucleotides and steroid hormones, and introduces a new concept of synthesizing both structural and functional characteristics of DNA at the same time within a single unified complex of interleaved tetra-ringed structures, steroid molecules which form reaction vessels that serve as co-enzymatic building blocks. The new theory indicates that the establishment of the DNA nucleotide code is among the very first synthesis steps.  Moreover, as a consequence of the intermolecular synthesis of both structural and functional characteristics within a unified complex, there is a culminating process step that sets forth in motion both DNA and the steroid structures that subsequently trigger replication and transcription, as well as protein translation, thereby resulting in the instantaneous release of life function.  
\end{abstract}

\newpage

\section*{Graphical Abstract}
\begin{figure}[!htb]
\centering
\includegraphics[width=.99\textwidth]{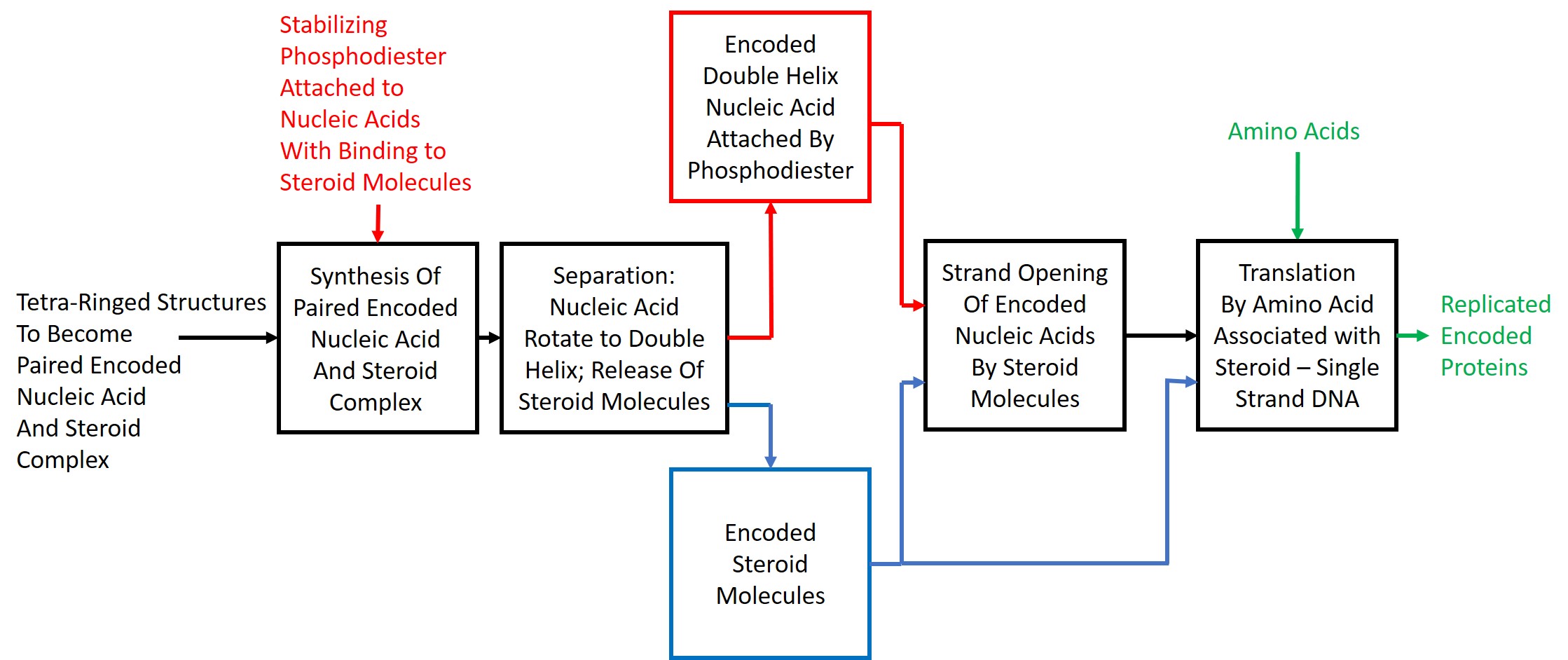}
\caption*{A schematic of the methodology to produce encoded proteins starting from the basic building blocks of the two steroid molecules.  A complex is formed in which the blocks are configured and arranged to establish enzymatic environments to induce amination reactions to co-synthesize within a unified complex DNA nucleotides and steroid molecules that ultimately, upon release, control transcriptional and translational function. To complete the complex, the reaction vessels are wired together along the edges through phosphodiester covalent attachment.  During this encoding process, an event in line with the second law of thermodynamics enables the separation of the encoded DNA structure into a double helix, and an associated set of steroid molecules that enable function.  The steroid molecule may then bind onto the DNA strands, which will cause its separation.  The new access will enable the encoded steroid structure to act as an intermediary and induce protein manufacturing through assembly of amino acids in a nucleotide triplet format organized through binding with the corresponding steroid molecules.   }

\end{figure}

\clearpage
\newpage
\section{Introduction} \label{sec:intro}

The origin of life is inherent to central tenets of science and philosophy; thus, as it remains one of the greatest unsolved mysteries, many prospective theories have been developed and studied at great length \cite{dyson1999origins, lazcano1996origin, sutherland2017opinion}. To focus this broad topic, as all cellular organisms of life contain DNA, the origin of life is synonymous with revealing the mechanisms responsible for synthesizing DNA starting from basic materials and processes that would have come into existence about the time Earth was formed, over four billion years ago \cite{cojocaru2017origin, filee2000origin}. In terms of the creation of the first DNA molecule, there are excellent articles written on the range of ideas, such as those that revolve around starting with RNA molecules and others that are based upon geological events \cite{walter1986origin, goodman1962formation, miller1988submarine, nakashima2018geochemistry}. Thus, in putting forth a new theory on the origin of DNA, as this article develops, it would be distractive to compare the new ideas directly with all of the alternative theories, because of the wide range, and varying levels of evidentiary basis \cite{lanier2017origin,hoshika2019hachimoji}. Therefore, in this introductory section, the objective is to characterize the constituents of a valid theory that reveals the synthesis of life function. For each requirement, the results of this article are highlighted that characterize a new theory on the origin of DNA.

\subsection{Requirements}

In developing a theory for the origin of life, a difficult problem to address immediately is the so-called ``chicken and egg dilemma”, as to which one came first \cite{service2015}.  For example, in the case of the transcription process under current theories, DNA is required to form proteins, but proteins are needed by DNA to initiate transcription.  Thus, there has been the speculation that DNA came about from an ``RNA world”, and this theory has taken the preeminent position in explaining the origin of life \cite{patel2015common}.  Thus, this approach would generate DNA from building blocks, but it would be unclear as to the driving force behind this self-assemble process of significant complexity.   In the theory presented in this research article, it is shown that the answer to the question of which one came first, the chicken or the egg, the answer is both:  that is, the DNA molecule and the steroid molecules that trigger replication and transcription were developed in an intermolecular linked manner at the same time in tandem with the steroid molecule affecting the DNA molecule and vice versa, both starting from the same molecule type, paired together and co-synthesized in a unified enzymatic complex.

A valid theory on the origin of DNA should also provide answers to basic questions on  its molecular structure.  For example, the following questions should be answered: why are purine and pyrimidine the bases that comprise DNA?  How are the arc distances between pairs of nucleotides determined?  Why does adenine-thymine only have two internal hydrogen bonds, whereas cytosine-guanine have three internal hydrogen bonds?  How is the synthesis of DNA related to the formation of the double helix?  What is the structural basis for DNA nucleotides?  Many theories on the origin of life never even get to this level of detail to provide a reasonable response to these basic questions, and most suggest that it is this way, because it has to be this way.  In this article, however, a process sequence is developed that within its first few steps provides quantitative reasoning to such structural questions on DNA, as well as its initial prebiotic functional application of protein translation.   

Another important requirement in the development of a theory on the origin of DNA is that the materials and processes utilized in forming DNA should be feasible during the time when  Earth was formed.  Some have considered DNA to have been possibly constructed up to 4 billion years ago, which is close to the estimates on the time at which Earth was formed \cite{ preiner2020future}.  To accommodate for further pushbacks, the theories should even consider DNA to be generated essentially at the same time as the formation of the earth.  Thus, lengthy evolutionary processes, or theories based upon chance of multiple sequential events that have low yields, would be hard to satisfy this category.  The materials and processes should be developed that are consistent with these conditions, and the processes should be relatively rapid, and preferably simple.  Most theories on the origin of DNA are valid in applying chemicals and processes that would have been present at the time when Earth was formed, but have challenges in terms of its simplicity.  However, as shown here, the materials in the proposed theory and process steps are consistent with these conditions, and the processes are of an elementary nature defined  at the onset of the synthesis.

Ideally, the new theory on the origin of DNA should not only explain the past, but offer a path going forward.  Approaches which utilize actions by chance or require lengthy development periods are less interesting than those that can be used in synthesis applications or to discover new medical knowledge.  DNA is central to the operation, and a better understanding of it should result in new medicines, new therapies, and new approaches.  Random events are thus not satisfying in the development of a theory for the creation of life function through DNA, and are not helpful in addressing basic human inquiries on the origins of life.  In this article, it is shown that DNA  is the result from the synthesis of a logical set of steps, which can lead not only to a better understanding of DNA, but also to produce new ideas such as in the transcription process itself.

One great challenge in the development of a theory on the origin of DNA is in testing theory, which may imply the capability to recreate it.  That is a daunting challenge as for some theories it is not possible to develop a protocol to test the hypothesis, especially those that rely on slow periods of evolution, random events, or exogenous processes.    However, if it were possible to have a ``chemical fossil" of DNA, which has underlying characteristic of the structure of DNA, then it would be possible to have a first line of comparative analysis.   In this article the identification of a discovery of a chemical fossil of DNA is identified for the first time, in which it is indicated that the molecular structure of the steroid molecule is embedded into each and every DNA nucleotide to the present day.  Thus, the theory developed in this article is matched to this discovery, thereby providing a verifiable metric of its validity, which the alternative theories do not address.   Moreover, it is further shown through analysis of measurable efficacy of certain pharmaceuticals, that the results are consistent with the structural arrangements involved in the molecules comprising the strategy of synthesizing DNA and steroid molecules by a basic set of processes, and hence provide a biomarker as evidence of the developed theory.

Most importantly and generally the largest stumbling block is that the new theory has to resolve the origins of the ``code of life”.  Literature acknowledges that this is the biggest problem for the current theories \cite{kun2018evolution, koonin2017origin}.  There is no clear-cut answer, or even anything close in explaining the code of life as AGTC, including those that start within the RNA world.  Thus, while the theories describe how the molecules can be formed, they do not characterize the organizational strategy associated with the capability of DNA to store and retrieve information.  In the second step of the proposed process flow, it is shown how the code of life is generated and its functional nature established.  Furthermore, the methodology enables the mapping of the base 3 nucleotide triplet to amino acid via the steroid molecule, described in the book \cite{schaper2020design}. Combined with the alignment of the results with experimental evidence including those consistent with the fundamental molecular structure of DNA, these are significant reasons to support the validity of the developed theory on the origin of DNA.  It is the only one that can explain the communication channel and orgin of information content, along with the structure and function requirements for life \cite{schaper2020design}.

Finally, a new theory on the origin of DNA should have a novelty, that is an insight that brings something new to the table.  In this research article, the concept of synthesizing the paired mechanisms of the structure and function of DNA simultaneously in a unified complex, in which structure and function are effectively designed as one.  Thus, when the co-formative processes are complete, DNA and its associated steroid molecules  are put into practice immediately to establish in an instant a central life function of replication and protein translation through a transcription procedure.  Therefore, the equivalent of a big bang moment for life function is enabled, of which there is paired synthesis of DNA and steroid molecules, developed in tandem as a co-synthesized intermolecular unified complex, dependent upon each other in one sequence of steps, co-optimized in structure and function, released simultaneously.  

Furthermore, since the purpose of DNA is to translate to proteins, a valid theory on its origin needs to describe the mechanism by which the complementary base pairs map to amino acids.  Again, almost all alternative theories do not address this issue as they do not even have the capability to determine how the base pair code is defined in the first place, and thus leave this important component unattended.  However, in the theory described in this article, this mapping of DNA base pair to amino acid is achieved for constructs, and importantly, indicate that it is a natural outcome of the procedure of associating the integrated, paired synthesis also permits encoding of an intermediary to achieve the amino acid to base mapping, which is fully described in my recent book that follows from this work \cite{schaper2020design}.

Although there are challenges in developing a theory on the creation of DNA, it is possible to develop a checklist for the requirements associated with an explanation on the origin of life.  In so doing, exciting explanations of the past are possible, unraveling many mysteries, and a path going forward is developed for exciting new discoveries.  Therefore, as this theory that will be described herein satisfies this checklist of basic requirements, it should enable this theory as a leading candidate among all theories describing the synthesis of the structure and function of DNA, its molecular triggers, that is specialized molecules of a steroid construct, for protein production,  and hence enable a chemical origin of life.

\subsection{Historical Context}
As this new theory will be disruptive and is quite distinct to the well published approach associated with the RNA world among other approaches, it can have appearances of ``fringe science", at least according \cite{hooft2015}.  If this may be the case, it is recommended that in order to introduce the new theory, a comparison back to the original concepts is warranted, in which the current mainstream approach can be compared against the proposed new approach.  Thus, in addition to the checklist of Section \ref{sec:intro}, an apples-to-apples comparison can be made as to analyzing the take-off point of current approach versus the new point, pertaining to the question:  what new insight is there that requires a new way of thinking, in this case about origin of DNA.

In this regard, an introductory paragraph of \cite{filee2000origin}, while brief, is informative.  Referencing the book by Monod \cite{monod1974chance}, it is noted that molecular biologists, in the years hitherto the discovery of DNA structure to the modern approach to study of its origin, considered life function to coincide with the synthesis of the first DNA molecule.  Moreover, they noted that Watson and Crick speculated of a quality of DNA which permitted protein generation without the need for proteins, quoting: ``whether a special enzyme would be required to carry out the polymerization or whether the existing single helical chain could act effectively as an enzyme” \cite{watson1953molecular}.  Further, \cite{filee2000origin} note that this thinking was in line with Schrodinger’s prediction of an aperiodic crystal as characteristic of life function \cite{schrodinger1992life}.  The research community however did not take this direction, however and opted for the current linear logic of independently synthesized molecules of RNA leading up to DNA.

\subsection{Advancements and Consistencies}

In this research article, the new theory and developments are aligned with the speculation of Watson and Crick, Schrodinger's aperiodic crystal, and the original molecular biologists, who favored DNA as the origin of life;  but adds a new concept to enable reality for this line of thinking:  Here, I aim to prove that  the original DNA molecule was but half of a co-enzymatic, co-synthesized, co-encoded ``master" complex, in which the other half was comprised of ultimately its triggering elements, that is steroid molecules, which enabled life function persuant to DNA when both were simultaneously released in a decomposition of the master complex.  The details will follow on how this fits together, and although it may seem complicated, is actually simple and is pursuant to Occam's Razor, in that as soon as the master complex is synthesized, and separated, life function is created, and the evolutionary processes can begin, that is, the originating work is done with sufficient concentration of the released molecules to permit further development immediately.  Moreover, this approach will satisfy among others, one important conundrum that has puzzled theoreticians: the second law of thermodynamics related to the ordered nature of DNA.  In that DNA is derived, along with its triggering elements, it is the disordering of the master complex into two ordered complexes that explains the satisfaction of the second law, or the tendency to induce disorder from order, where DNA becomes the product of a disordering event, as will be seen, to lower the overall energy of the composite system.

The new results are described in my new book \cite{schaper2020design}, whose contents are outlined in Appendix \ref{app:book}.  The results include the communication channels of information theoretic starting from encoding of the unified complex, transmission in the form of a double helix and steroid molecules, and then access of the double helix by the steroid molecules, so as to produce a chain of amino acids forming a protein molecule.

\section{Results}

To develop the new theory, experimental data is first described in support of it, thereby indicating a discovery that I had made in \cite{schaper2020endogenous, schaper2020structural} that there is a common structure, the steroid molecule, embedded in each in every DNA nucleotide pair, which will serve as a beacon to develop the new theory on the origin of DNA.   In addition, it is shown that this can be extended to match steroid hormones structural to DNA, and even indicate functionality as a present-day pharmacological biomarker associated with a process that can be correlated with the structure of DNA.  After presentation of the experimental evidence, which serves as the basis for the theoretical developments, the process steps are described to synthesize not only encoded DNA, but also encoded steroid molecules.  These units are co-developed at the same time within the same unified complex, and when released can be put into practice immediately.  For prebiotic conditions, it can serve as the basis to generate the protein structures that can later take over in function as known today, such as the enzymes involved in transcription and translation.   The appendix describes reaction mechanisms to support the process steps indicated in this section.

\subsection{Experimental Evidence}
Three levels of experimental evidence are indicated to support the theoretical developments.  The first experimental evidence is that of a chemical fossil of DNA in which there is a basis molecule, that is a structural basis, embedded within each nucleotide pairing of DNA, which consists of a four ringed steroid molecule.  The second experimental evidence is a molecular fossil associated with DNA as the alignment of steroid hormones to the DNA nucleotide pairs, which is important in the paired development of both molecules within the same complex.  The third experimental evidence is pharmacological information as a biomarker of DNA indicating the importance of the paired coupling of DNA nucleotides and steroid molecules.

\subsubsection{Embedded steroid structure as a chemical fossil of DNA}
\begin{figure}[!htb]
\centering \hspace{2cm}
\subfloat[Front]{\includegraphics[width=.35\textwidth]{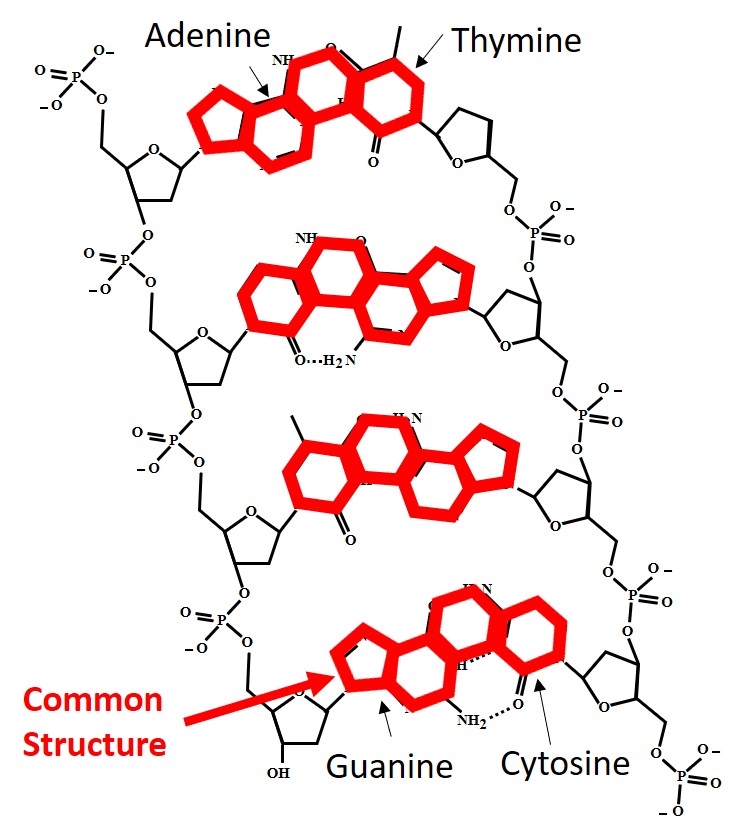}} \hfill
\subfloat[Back]{\includegraphics[width=.35\textwidth]{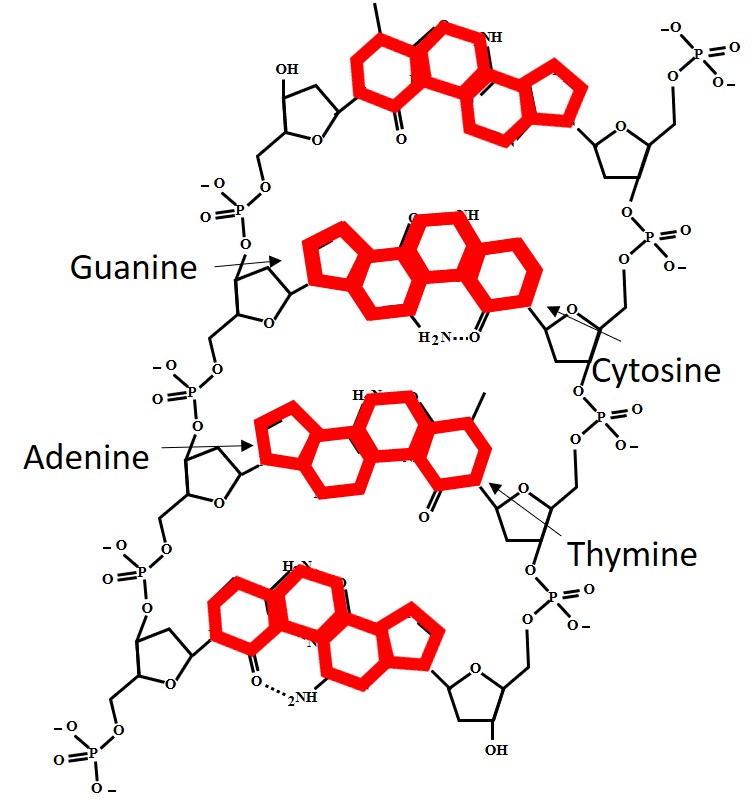}} \hfill
\caption{\label{fig:dna}   (a) In a structural analysis of the four possible DNA nucleotide pairs presented from top to bottom as Adenine-Thymine (A-T), Cytosine-Guanine (C-G), Thymine-Adenine (T-A), and Guanine-Cytosine (G-C), there is a common structure in the shape of a four ring steroid molecule at each of the nucleotides.; (b) Examining the ``back" side of the normal presentation, which is  equivalent to that which would be seen on the back of a printed page with the molecule on the front, the common structure of the steroid molecule is obtained for each of the nucleotide pairs.  The coupling of back side to front side is achieved by flipping the common structure, thus indicative that one common structure is appropriate for each of the pairs as the backside orientation becomes reoriented properly on the frontside, and vice versa.  
}
\end{figure}

To develop an experimental basis to validate the theory, a structural analysis of DNA nucleotide base pairings is analyzed.  In Figure \ref{fig:dna}(a), the front of the DNA nucleotides is indicated for each of the four pairings as listed from top to bottom in the figure as Adenine-Thymine (A-T), Cytosine-Guanine (C-G), Thymine-Adenine (T-A), and Guanine-Cytosine (G-C).  In Figure \ref{fig:dna}(b), the backside of each of the four DNA nucleotide pairings is illustrated.  This is obtained by flipping the front orientation in a horizontal direction, and both sides are important to the analysis.  Also included in the illustrations are the phosphodiester backbone and the sugar molecules.  The arrangements are envisioned as the DNA nucleotide structure printed on paper, and viewing it from the front and the back side.   Also included in the illustration is the hydrogen bonding of the base pairings, in which it is shown that there are three hydrogen bonds for the C-G and G-C pairs, and two internal hydrogen bonds for the A-T and T-A pairs.  It is of importance that at the carbon two position of thymine, there exists a functional oxygen element, but there is no corresponding element on adenine that could provide a hydrogen bond as a link.  

To indicate that the four fused ring structure of a steroid is a ``chemical fossil of DNA", the extraordinary result is presented in the four red overlays each of Figure \ref{fig:dna}(a) and (b).  This overlay is obtained by connecting precisely a total of seventeen oxygen, nitrogen, and carbon elements of the base pairs in a particular manner to produce a tetra-ring structure embedded within each of the four DNA nucleotide pairs.  This can be achieved for each base pair, front and back, and the match is in perfect correspondence when comparing each of the matches.  The four-ring structure is exactly in the shape of a steroid molecule, which was originally identified in the preprints \cite{schaper2020endogenous, schaper2020structural}.  The orientation of the molecule is in perfect correspondence when analyzing the frontside for the A-T and G-C pairings, where a purine molecule starts the association on the 5' side, while on the backside the alignment is perfect for the T-A and C-G pairings when the pyrimidine molecule starts the 5' pair.  There is a stretching along one of the rings where the purine and pyrimidine come into alignment through hydrogen bonding.  Thus, this association is a chemical fossil of DNA of which theories on the origin of DNA must identify.  The theory developed in this research article begins from the steroid molecule, and thus is in perfect harmony with the basis defined by this chemical fossil of DNA.   

\subsubsection{Alignment with steroid hormones as experimental structural symmetry}

\begin{figure}[!htb]
\centering 
\subfloat[Front with Steroid Hormone]{\includegraphics[width=.3\textwidth]{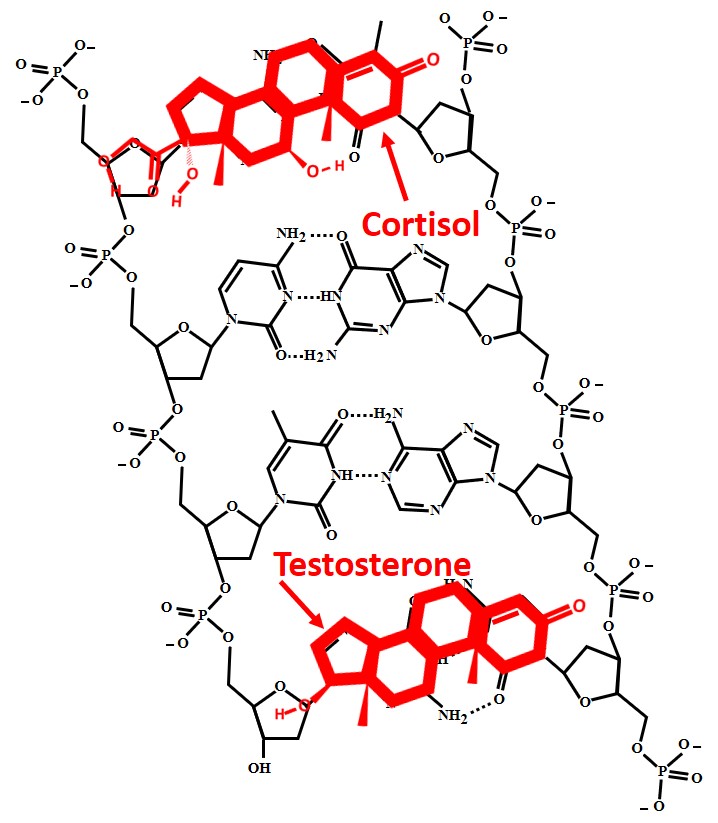}} \hfill
\subfloat[Back with Steroid Hormone]{\includegraphics[width=.3\textwidth]{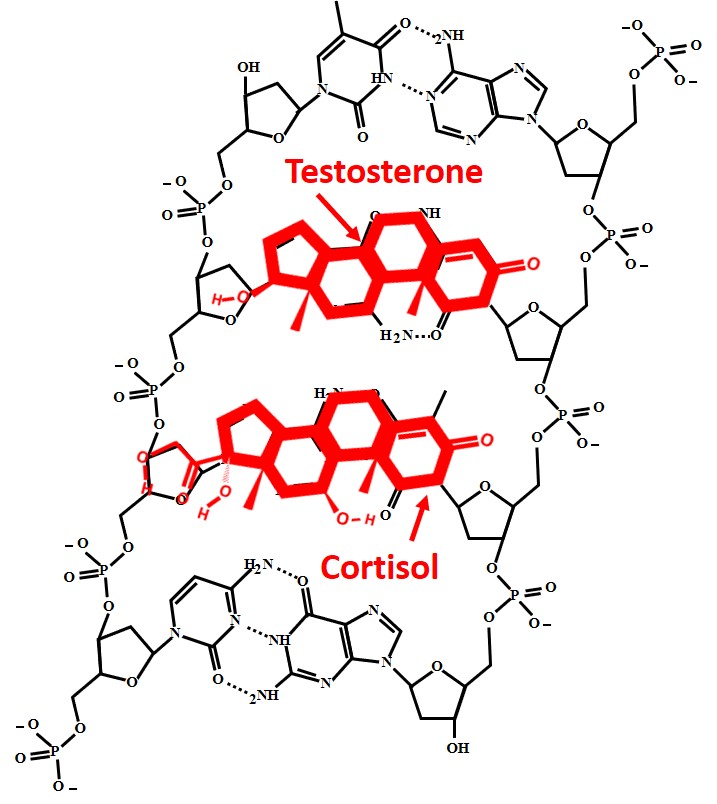}}  \hfill 
\subfloat[New Code for DNA Nucleotides]{\includegraphics[width=.35\textwidth]{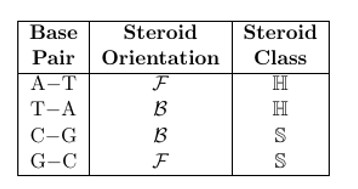}}  \hfill 
\caption{\label{fig:Mols} Consistent with the result of a common steroid molecule embedded in the DNA nucleotides, the intermolecular bonding capability is examined with respect to thymine.  Classes of steroid molecules are defined with cortisol defining those molecules which have a mid-molecule functional group, and testosterone representing those that do not have a mid-molecule functional group.  The objective is to overlay the steroid such that three hydrogen bonds are formed. on the common steroid structure for each of the four possible steroid structures, a class of steroid hormones are assigned with the objective of achieving a total of three hydrogen bonds when overlaid with the steroid hormone.  In (a), the cortisol steroid hormone forms an intermolecular bond with the (A-T) pairing, and thus a total of three hydrogen bonds result.  In the case for (G-C), there is already three internal hydrogen bonds, and thus testosterone is used as the overlay since it does not have a mid-molecule hydroxyl group.  For cortisol and testosterone, the end groups are available for connection with the phosphodiester backbone, which was shown in the simulation.  (b) Following the same protocol, cortisol forms an intermolecular hydrogen bond with the (T-A) pair, while testosterone couples with the (C-G) nucleotide pair, but on the back-side.  This is consistent with the need to rotate the pyrimidine molecules at the 5' strand to achieve the double helix when ejecting the steroid triggering molecules. (c)  The  location of the steroid structure is indicated as, ${\mathcal{F}}$ for front and ${\mathcal{B}}$ for back, and the corresponding class of structurally compatible steroid class, ${\mathbb{H}}$ indicates the class of cortisol-like molecules, which has a functional group positioned to interact with the available functional group of thymine, and ${\mathbb{S}}$ signifies the class of testosterone-like molecules, which do not have a functional group positioned that can interact with the available functional group of thymine.}
\end{figure}

The second line of support involves the symmetry of the steroid hormone with the DNA nucleotide pairs with respect to the intermolecular binding.  Namely the corticosteroids, which have a mid-molecule group, should bind onto thymine.  This result is consistent with the hydrogen bond that was presented very early into the development, essentially the definition of the DNA nucleotide code itself.  It should be the case that the symmetry of the steroid molecule is capable of forming an intermolecular hydrogen bond with A-T and T-A, and not needed for C-G and G-C, the result is evaluated with the representation of DNA and the steroid hormones.  As shown in Figures \ref{fig:Mols}(a) and \ref{fig:Mols}(b), this symmetry is indicated and shown.  Of course, having developed the symmetry relations between steroid hormones and DNA nucleotides as described in the preprint \cite{schaper2020structural}, this mapping was an objective.  Thus, it is apparent that this representation is still conserved to this date, which includes the relationship with steroid hormones, the orientation, the intermolecular bonds, the connection to the phosphodiester backbone, and even the configuration of the end groups.  Moreover, together, the orientation and class of steroid hormone correspondence form a new code for DNA as shown in Figure \ref{fig:Mols}(c), which is further described in the preprint \cite{schaper2020new}.  Moreover,  this is consistent with the indication in Section \ref{sec:resultsfirst} which noted that the nucleotide pairs with pyrimidines along left (5') side had to be flipped in order to achieve their proper orientation consistent with the double helix.  Hence it is logical that the steroid representation would involve a forward side and a reverse side of orientation.    This configuration of forward and reverse orientation is thus apparently still conserved to this date  in the DNA molecule.   Thus, the second line of experimental evidence of a structural biology nature to identify a ``molecular fossil" of DNA, which is the structural match of the steroid hormone, is in support of the new theory on the origin of life is successful of which steroid molecules and DNA nucleotides are synthesized together within the same unified complex. 

\subsubsection{Pharmacological efficacy of steroid hormones as a biomarker of DNA functionality}
\begin{figure}[!htb]
\centering
\hspace{20mm}
\subfloat[Prednisolone]{\includegraphics[width=.35\textwidth]{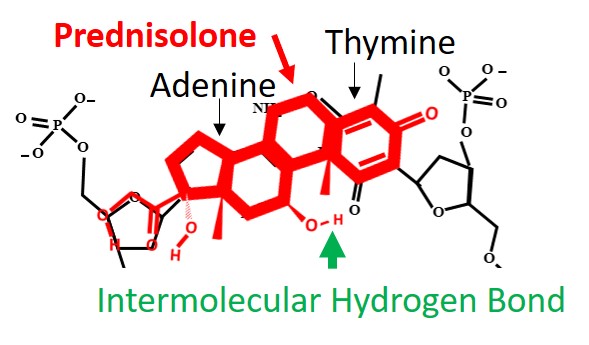}} \hfill
\subfloat[Prednisone]{\includegraphics[width=.35\textwidth]{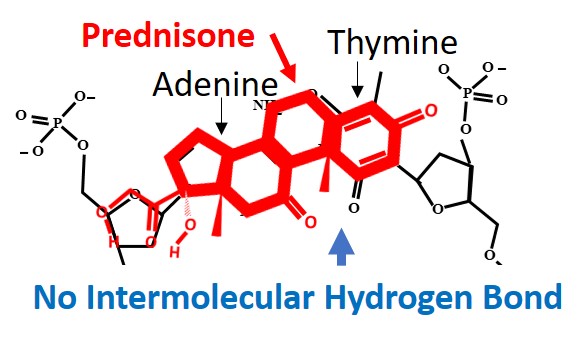}} \hfill \\
\hspace{20mm}
\subfloat[Cortisol]{\includegraphics[width=.35\textwidth]{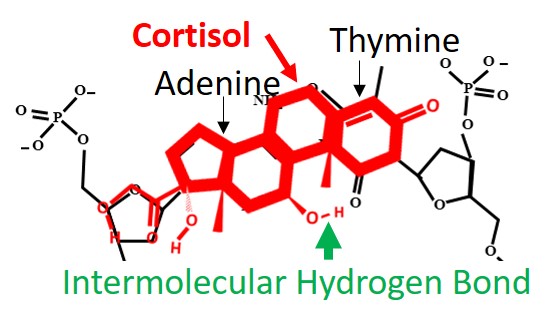}} \hfill
\subfloat[Dexamethasone]{\includegraphics[width=.35\textwidth]{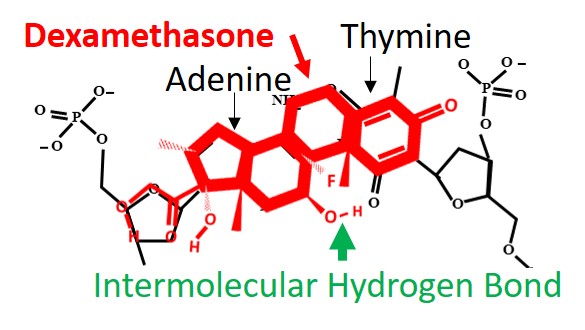}} \hfill
\caption{\label{fig:Roids} For analysis of experimental results, the efficacy of steroid therapeutics is examined.  (a) The capability to form an intermolecular hydrogen bond between prednisolone and thymine.  (b)  It is not possible to form an intermolecular hydrogen bond between prednisone and thymine.  Indeed prednisone is inactive in practice, and must be converted to prednisolone in order for it to be potent.  This result provides evidence in proving the claims of this manuscript.  (c) The overlay of the nominal steroid hormone cortisol is shown capability to form an intermolecular hydrogen bond with thymine.  With respect to prednisolone, the double bond in the ring adjacent to thymine aid in enhancing the intermolecular hydrogen bond strength, and thus increased potency is expected for prednisolone relative to cortisol, which is indeed the case in practice.  (d)  Dexamethasone also is capable of forming an intermolecular hydrogen bond with thymine.  In addition, the fluoride group and double bond will aid in the potency, and thus it would be expected to have greater potency than prednisolone, and cortisol, which again aligns with that observed in practice}
\end{figure}

To further validate the method, experimental data is evaluated that can be used to prove the claim of this theory that the intermolecular bond which essentially established the DNA code is critical, and thus serves as a biomarker of DNA to guide in the development of a theory of its origin.  In this experimental analysis, the intermolecular hydrogen bond is examined of coupling steroid hormones similar to cortisol to thymine for corticosteroid data.  The handbooks indicate   that the relative activity of the steroids  dexamethasone:prednisolone:cortisol is approximately 25:4:1 and that prednisone, in its native form, is inactive as it must be converted into prednisolone in the liver before it can be used \cite{becker2013basic}.  This relationship is consistent with the intermolecular bond to thymine by its steroid triggering molecule as presented in this study. In Figure \ref{fig:Roids}(a), the relationship of prednisolone to thymine is indicated, in which it is shown that it is possible to form an intermolecular hydrogen bond with thymine.  However, in Figure \ref{fig:Roids}(b), the relationship of prednisone to thymine is indicated, in which it is not possible to form an intermolecular hydrogen bond.  Note that the OH group is the only difference between prednisolone and prednisone.  This  precise result is consistent with the efficacy of prednisone, which experimentally shows no activity, and must be converted into prednisolone before becoming active.

Further experimental data is consistent with the theory.  In Figure \ref{fig:Roids}(c), the relationship between the steroid hormone cortisol and thymine is indicated, which can form an intermolecular hydrogen bond.  With respect to prednisolone, the only difference is an extra double bond in the ring of prednisolone within thymine.  The additional double bond will aid in stabilizing the intermolecular hydrogen bond of thymine and prednisolone, and may also provide favorable structural arrangement in terms of the angle.  Further, by examination of dexamethasone in Figure \ref{fig:Roids}(d), the activity will be even further enhanced, due to the fluoride group stabilizing the intermolecular hydrogen bond even more than prednisolone, thereby increasing its relative activity.  Thus, the third line of evidence based on experimental data is consistent with the paired development of the steroid molecule and DNA nucleotides.

\subsection{Unified Complex for Paired Encoded Nucleic Acids and Steroid Molecules} \label{sec:resultsfirst}

To derive the integration of structure and function within one unified complex, the developments begin from the work described in these \cite{schaper2020endogenous,schaper2020structural}, which  indicated the structural symmetry of steroid molecules, steroid hormones and DNA nucleotides.  Thus, to continue the developments, these molecules will be stacked and hence hydrogen bonding will stabilize the structure, as indicated in Figure \ref{fig:sterstack}.  This particular format is described later in the research article, Section \ref{sec:ch}, on how the diketone group, hydroxyl group, and five carbon element ring are formed from a  starting material.   In stacking these steroid molecules, intermolecular hydrogen bonding is deployed to secure the structure as indicated in Figure \ref{fig:sterstack}, noting that the rings are aligned, with the five carbon ring situated in alignment with the six carbon ring.  It is noted that these paired molecules are enantiomers, mirror images of each other.

\begin{figure}[!htb]
\centering
\includegraphics[width=.4\textwidth]{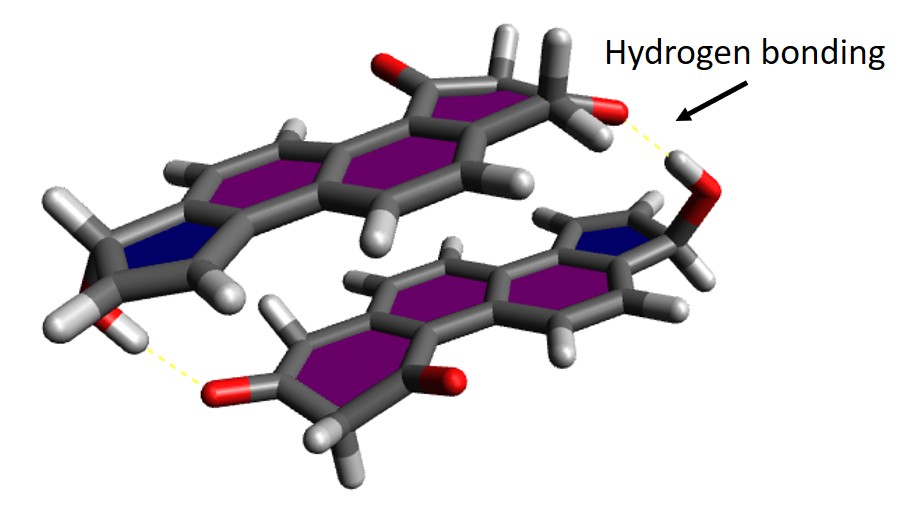}
\caption{\label{fig:sterstack} The developments focus on a paired structure of two molecules comprised of four fused rings, in the shape of a steroid molecule, with three six carbon aromatic rings, and one five carbon aromatic ring, that is functionalized at each end, one by a hydroxyl group and the other by ketones.  Hydrogen bonding secures the two molecules.  The generation of this arrangement is discussed later in the article.}
\end{figure}

\subsubsection{Enzymatic Coupling} \label{sec:enzyme}
As noted in reference \cite{filee2000origin}, in the original work of Watson and Crick speculated that the double helix of nucleic acids may have an enzymatic nature; but that idea did not take hold and since then the research direction has trended toward the linear logic of a synthesis starting from single-strand RNA nucleotides, which would ultimately lead to the DNA double helix complex.  However, based upon the analysis presented in this research article, it is possible to re-evaluate the speculation of Watson and Crick within an analytic framework of the origination of DNA as an enzymatic construct of interleaved tetra-ringed structures, of which one section is covalently bonded to a phosphodiester backbone, while the adjacent structure is connected through hydrogen bonding in close proximity.

As indicated in Figure \ref{fig:enzyme}, the coordination of the paired building blocks enables a sequence of effective reaction vessels, secured at the ends through phosphodiester covalent coupling.   This confined environment enables the potential for an enzymatic capability in which the temporary formation of bonding arrangements to permit intermolecular cyclic structures that can establish element substitution, the degradation of chemical bonds to reduce intramolecular stress, and the  differentiation of the templates into their individual capacity as a nucleotide base.  While the structures did have aromatic stabilization, the intermolecular structures may also have benefitted from resonance stabilization, and thus the intermolecular constraints to maintain an in-plane structure during the synthesis steps would have promoted the present-day configuration in which the original aromatic structure was traded for an amine based self-assembled construct of a planar configuration.   Relevant reaction sequences to this synthesis route are the stereoselective  Diels-Alder reactions \cite{siegel2010computational,sanyal2000stereoselective,wheeler2010probing}, for this type of integrated structure.

\begin{figure}[!htb]
\centering
\includegraphics[width=.65\textwidth]{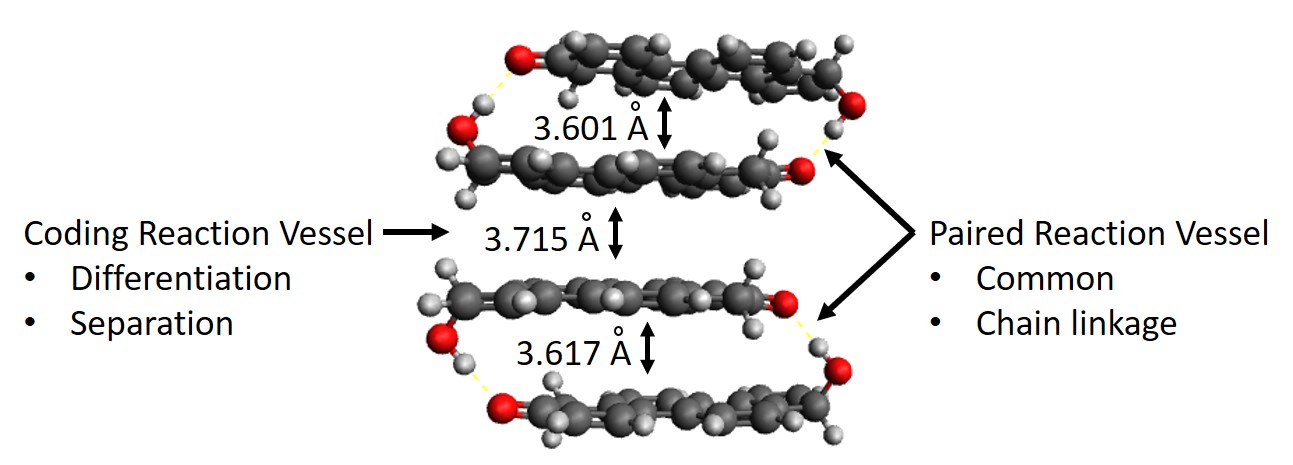}
\caption{\label{fig:enzyme} By interleaving the tetra-ringed aromatic structures, and securing each alternating pair to a phosphodiester backbone, there exists the capacity for an enzymatic function in the development of each molecule.  It is proposed that this constitutes a series of reaction vessels, which together with aggresive environmental conditions, the co-synthesis of the DNA nucleotides, and its triggering elements, comprised as steroid molecules, would have resulted from interactive co-development.  There are two styles of enzymatic reaction vessels.  The directly coupled are used in the synthesis of the coupling to the phosphodiester base, and the vessel that is defined by the meeting of two paired units, is used in the differentiation and separation of the DNA nucleotides and the steroid molecules.}
\end{figure}

\subsubsection{Encoding} \label{sec:encoding}

\begin{figure}[!htb]
\centering
\subfloat[A-T motif]{\includegraphics[width=.7\textwidth]{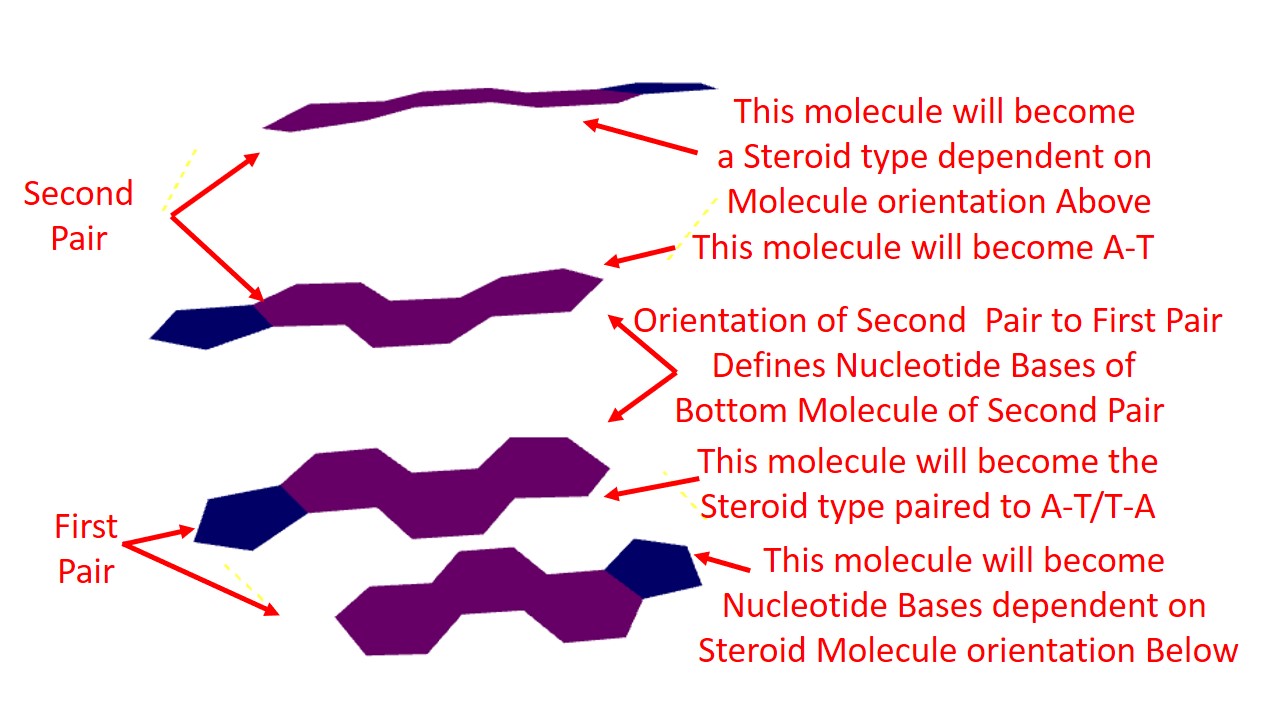}} \hfill\\
\subfloat[T-A motif]{\includegraphics[width=.32\textwidth]{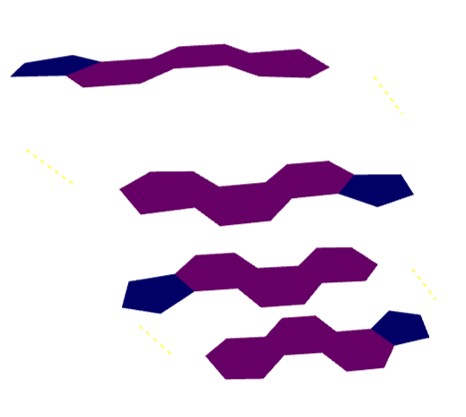}} \hfill 
\subfloat[G-C motif]{\includegraphics[width=.32\textwidth]{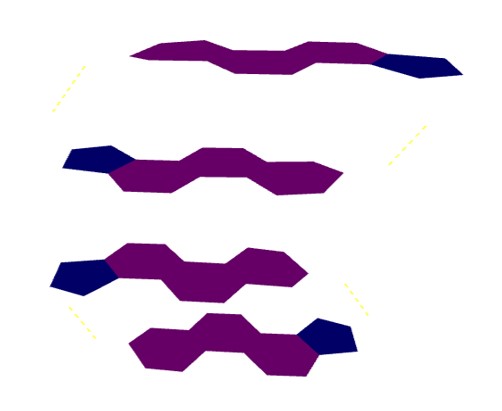}} \hfill
\subfloat[C-G motif]{\includegraphics[width=.32\textwidth]{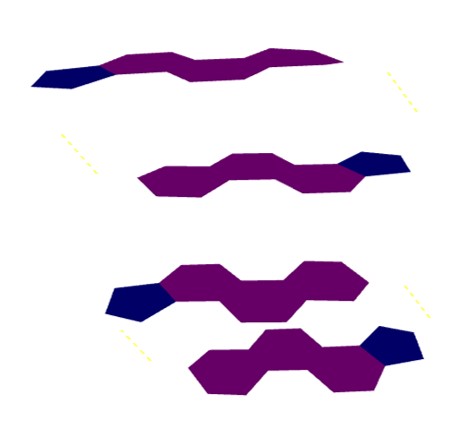}} \hfill
\caption{\label{fig:sterletter}As a major result for this research article, the code of DNA is formed by stacking two coupled, functionalized steroid molecules, in which here the DNA nucleotide code is defined for the first and third molecules, starting from the bottom.  In the representation, the configuration is shown for the third molecule, numbering started from the bottom, which after going through the steps described  in this article will result in the following nucleotide pairs:  (a) adenine-thymine (A-T).  (b) thymine-adenine (T-A). (c) guanine-cytosine (G-C).  (d) cytosine-guanine (C-G). In addition to the four types of base pairings of the DNA molecule, the paired steroid molecule will also be encoded into two types. }
\end{figure}

Now, the key to explain the encoding of DNA nucleotides is shown.  In stacking the next pair of steroid molecules on top of the initiating  pair of steroid molecules, the orientation will define ultimately the ``letters" of the nucleotide pair, whether it is adenine-thymine, thymine-adenine, guanine-cytosine, or cytosine-guanine!  There are exactly four configurations, when placing the second set on top of the first set, and as will be seen later, one molecule of the pair will define the nucleotide pair, and the other will define the cooperative steroid molecule that will interact with DNA to trigger replication and transcription.  For example, in Figure \ref{fig:sterletter}(a), the third molecule starting from the bottom will take on the DNA nucleotide pairing of adenine-thymine, after the process steps to be described  are completed.  By flipping the upper pair of molecules, that is molecules 3 and 4, as a unit vertically, the thymine-adenine base pairing will result as molecule 3 as indicated in Figure\ref{fig:sterletter}(b).  Rotating the upper pair of molecules as a unit one hundred eighty degrees will result in the guanine-cytosine base pair as molecule 3 as indicated in Figure \ref{fig:sterletter}(c), and flipping the upper pair of  molecules as a unit will result in cytosine-guanine as molecule 3 as in Figure \ref{fig:sterletter}(d).  It is noted that the determination of the nucleotide pair of molecule 3 is depending upon the starting orientation of molecule 1,  and thus any adjacent letter can be obtained independent of the starting point of molecule 1.  

\begin{figure}[!htb]
\centering
\subfloat[To become A-T]{\includegraphics[width=.4\textwidth]{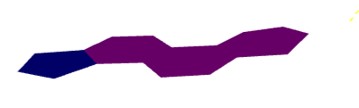}} \hfill
\subfloat[To become T-A]{\includegraphics[width=.4\textwidth]{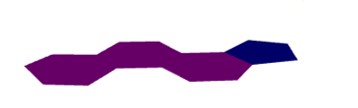}} \hfill \\ 
\subfloat[To become G-C]{\includegraphics[width=.4\textwidth]{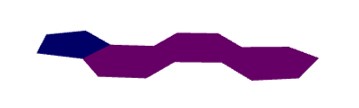}} \hfill
\subfloat[To become C-G]{\includegraphics[width=.4\textwidth]{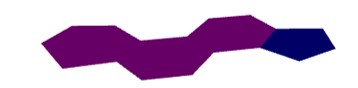}} \hfill\\
\subfloat[Referencing Paired Molecule to be Encoded Steroid]{\includegraphics[width=.4\textwidth]{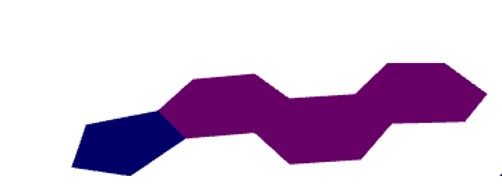}} \hfill
\caption{\label{fig:motif} The orientation of the adjacent molecule of the second pair with respect to the molecule of the first pair will define the final result. There are four orientations possible. In the example of Figure \ref{fig:sterletter}, the different orientations are shown.  As will be shown, the coupling elements between the adjacent molecules will be used to define the base pairing.  Note that there are four rings that define each structure, and that there are four and only four base pairings.  These molecules after paired synthesis with steroid structures will become (a) A-T, (b) T-A, (c) G-C, and (d) C-G.  (e) In addition, the paired steroid molecule will enable replication and transcription, which is the referencing molecule co-synthesized with the DNA nucleotides.  If rings 2 and 3 are aligned it will become a steroid interactive with (A-T) and (T-A); if not aligned, a steroid consistent with (G-C) and (C-G).}
\end{figure}

{\em Remarkably, that is all that is really necessary to create a coded DNA molecule as well as its steroid molecules, which will trigger genetic replication and transcription, that is the results of Figure \ref{fig:sterletter}.}  It is possible to create any set of letters from AGCT, as well as its steroid triggering components, by using the strategy of pairing and orientation, which is an amazingly simple process, as it needs to be!  Moreover, in addition to the base 4 code associated with the orientation of the pairings, another code is defined.  The steroid triggers have two orientations relative to the upper pair of molecules, meaning that there is alignment in terms of hydrogen bonding to either one side or another.  Thus, in addition to encoding what will become the nucleotide pair, it also encodes the type of steroid molecule, as affiliated with its hydrogen bonding location.  To further examine the orientation of the base-pairing, Figure \ref{fig:motif} indicates that there are four configurations, and only four configurations in planar relation to the underlying molecule, which will become its paired steroid.  It is also noted that in addition to coding the DNA nucleotides, it also encodes the steroid molecules.  As will be developed, the steroid molecules are coordinated with amino acids during the translation process, and thus will provide the specificity required for protein translation, as well as an inherent capability to induce strand separation.

\begin{figure}[!htb]
\centering
\subfloat[Example Rings as TAGTC]{\includegraphics[width=.5\textwidth]{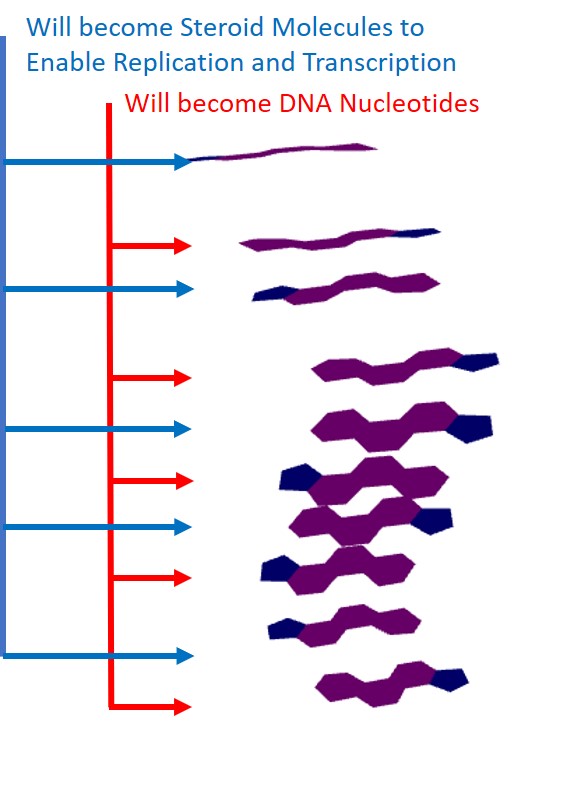}} \hfill
\subfloat[van der Waals spheres]{\includegraphics[width=.44\textwidth]{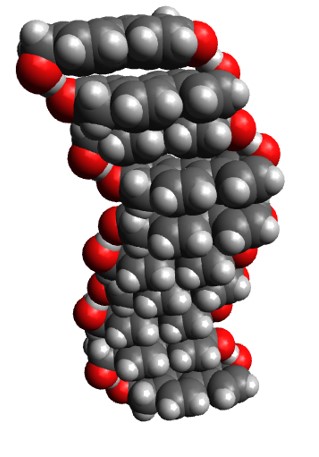}} \hfill
\caption{\label{fig:tenstack} (a) To use an example in describing the work, the structure is used which will result in the nucleotide pairs of (T-A), (A-T), (G-C), (T-A), (C-G) for molecules 1, 3, 5, 7, 9, numbering from the bottom.  It will also include four steroids configured as corticosteroids similar  to cortisol, for molecules 2 and 6, and similar to testosterone from molecules 4 and 8.  The tenth molecule would interact with the eleventh molecule.   (b) The van der Waals spheres indicate hydrogen bonding along two edges, and hydrogen bonding staggered  on either side of the structure.}
\end{figure}

To continue the discussion, it is interesting to use an example, such as the nucleotide representation of TAGTC.  In Figure \ref{fig:tenstack}(a), the stack of ten molecules for the creation of the five nucleotide pair of (T-A), (A-T), (G-C), (T-A), (C-G), starting from the bottom, is presented.  The other five molecules will be associated with steroid  molecules, which can be used for replication and transcription.  Both will be developed in the same complex, and both are interactive and influence each other.  The nucleotides will be formed on molecules 1, 3, 5, 7 and 9, and the steroids will be formed on molecules 2, 4, 6, and 8, with the numbering starting from the bottom.  There is another molecule below molecule 1 that influences its configuration, and would ultimately result in a steroid, but that molecule is paired, and in order to keep it brief, only the transformation on molecule 1 will be shown, and not the corresponding change to its paired molecule, which would be situated below it. Further, the starting orientation of molecule 1 is taken to indicate the independence of the letter sequencing.  The top molecule 10 only couples to molecule 9, and does not influence its configuration.  

\subsection{Strand Separation: Prebiotic Replication, Transcription and Translation} \label{sec:transcription}
Having produced the paired and encoded structures of both DNA and steroid molecules, the bases will only be accessible for replication and transcription if the strands are possible to separate.  Fortunately, the steroid molecule is available, and can bind onto the phosphodiester chain to enable strand separation.  This occurs through stabilization of the strands, which are vibrating and oscillating molecular constructs.  Thus there is kinetic energy transferred to potential energy through the stabilization process, which is an overall energy function whose internal energy must be dissipated through for example molecular disorganization, and bond breaking in this case.  An alternative would be an increase in temperature of the molecule, hence vibratory modes, but this may also induce strand separation.  Approximate mathematical representation for the process is indicated in Equations \ref{eq:a} and \ref{eq:b}, for the energy dissipation, $E_s(x)$ about the binding site, and the strand separation from its nominal position, $S(x)$.  The equations characterizing the strand separation are given by:
\begin{eqnarray} \label{eq:a}
E_{s}(x) &=& \frac{1}{2} (1 + \tanh(k_xx))\\ \label{eq:b}
S(x)&=&k_{b}(1 - (\tanh(k_xx) ^2))+S_{n}
\end{eqnarray}
where the parameters are associated with hydrogen bond energy $k_x$ = 1/40, coupling $k_{b}$=3, and nominal separation $ S_{n}$=1.8 {\AA}, and $x$ runs from 0 to 200 nucleotides, with the result than mirrored from -200 to 0 nucleotides about the binding point.   These equations can be used to approximate the strand opening away from the binding site.  It is dependent upon the energy associated with the hydrogen bonding keeping the strands intact, as when the term $\tanh(k_xx)$ approaches unity, the strand separation will be equal to its nominal value.  Other formulations for strand separation can use wave equations to describe the motion of the DNA strands, which will be constrained by the cross-coupling of the steroid molecule, and thus result in reflected wave patterns that will induce interference patterns, and bond separation for bubble formation.

\begin{figure}[!htb]
\centering
\subfloat[van der Waals spheres]{\includegraphics[width=.32\textwidth]{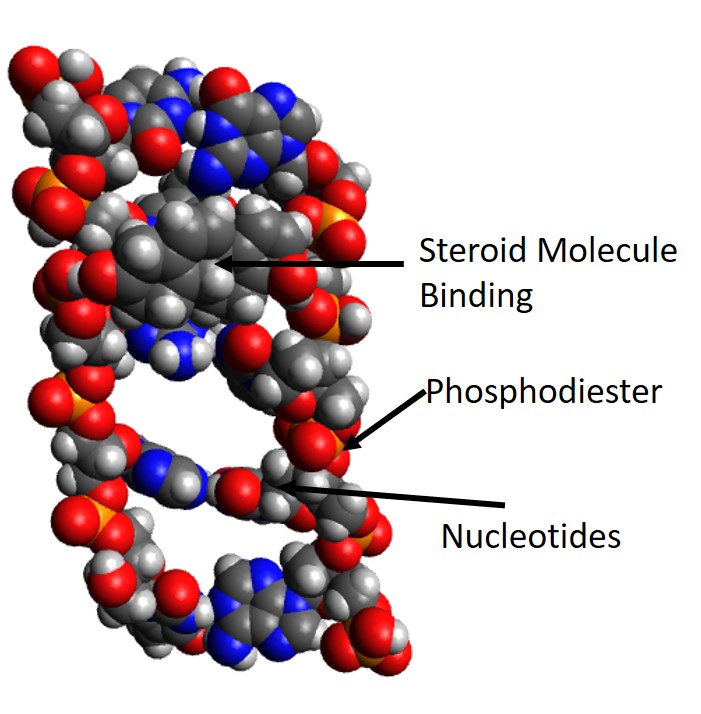}} \hfill
\subfloat[Electrostatic Potential]{\includegraphics[width=.2\textwidth]{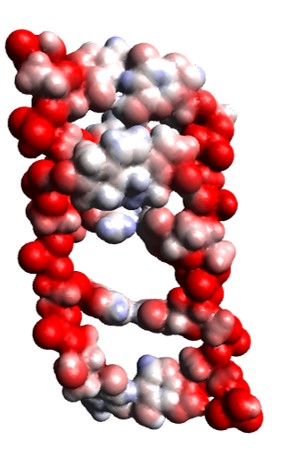}} \hfill
\subfloat[Strand Separation]{\includegraphics[width=0.35\textwidth]{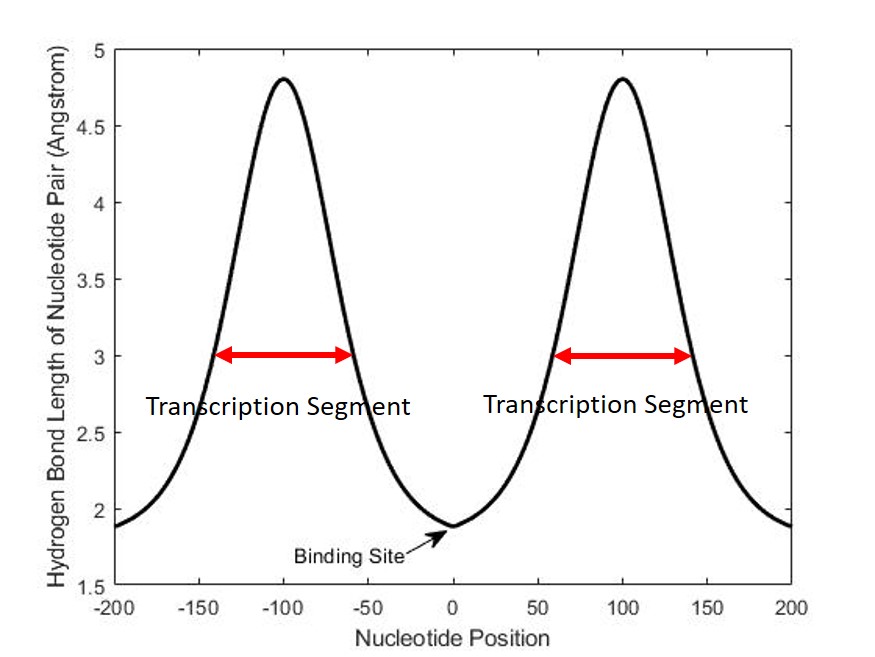}} \hfill 
\caption{\label{fig:DNAbind} (a) The binding of the steroid molecule onto the formed DNA complex to induce strand separation.  (b) The electrostatic potential of the binding site, which is achieved through hydrogen bonding at both ends, attaching to the phosphodiester coupling. (c) The formation of the transcription bubble on either side of the binding point up to two hundred nucleotides.  The quality of the binding point will determine the width of the transcribing separation of strands. }
\end{figure}

As the DNA molecule is stabilized by binding with the steroid molecule, there must be a balance in energy in terms of a release through bond breaking to be nearly equivalent to that of the stabilization provided by the binding event.  Essentially, the DNA strands act as two strings in constant motion, which is fluidic, and with the cross-bonding induced by the steroid molecule, that is the triggering element, it acts as a pinch-point, and the strands, which were in relatively independent motion become constrained at one point, and around that point, a small bubble must form through hydrogen bond breaking upstream and downstream of the pinch-point.  In Figure \ref{fig:DNAbind}(a) the binding of the steroid molecule onto the formed DNA complex is indicated in a molecular model as to its feasibility.  In Figure \ref{fig:DNAbind}(b) the electrostatic potential is indicated of the steroid molecule to show that the negative potential of the phosphodiester zone will enable alignment of the steroid molecule for binding.  In Figure \ref{fig:DNAbind}(c), the profile of the strand separation is approximated on either side of the binding site.  It is possible to deploy multiple steroid molecules to induce a stronger association and knoting of the DNA strands, and thereby improve the duration of the subsequent opening of the strands of the nucleic acid. 

\subsection{Translation for Protein Production}
 With this chemical procedure, the prebiotic production of encoded proteins originating from nucleic acids can be precisely defined, both in its initiation as a prebiotic function and transition to the present-day methodology, whereas, other than the present method, alternative theories on the origin of life have very little if anything to say about the methodology of protein translation.   The steroid molecule is a key component, as it enables the guidance of the DNA molecule, thereby encoding both itself and the DNA structure, and also effectively results in encoded amino acids, which can be mapped to the DNA sequence through the intermediary steroid molecule.  See \cite{schaper2020design} for more information on translation using the steroid molecule in combination with DNA to develop an interaction vessel of the base 3 sequence.  Moreover, the steroid molecule enables the strand separation of the DNA molecule to provide access to the nucleotide sequence.

\begin{figure}[!htb]
\centering
\includegraphics[width=.99\textwidth]{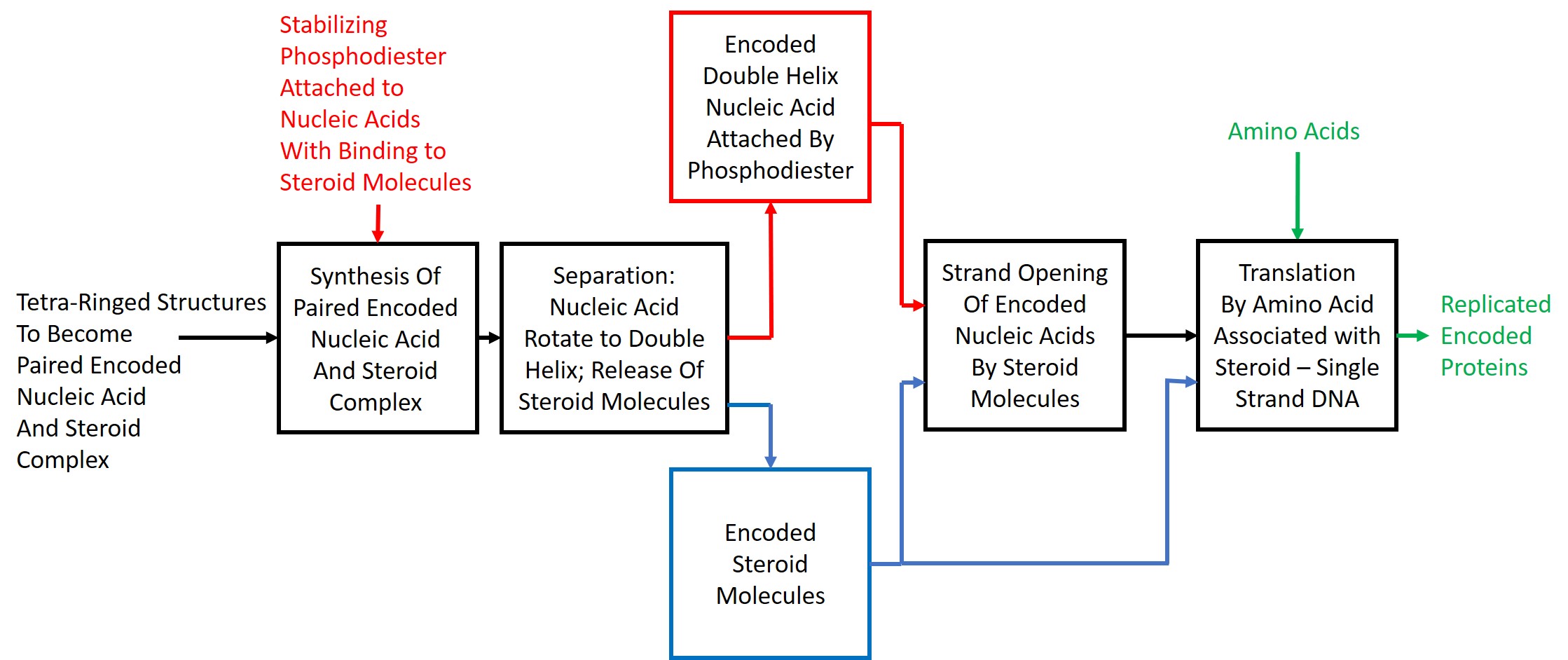}
\caption{\label{fig:translation} A schematic of the methodology to produce encoded proteins starting from the basic building blocks of the two steroid molecules.  A complex is formed in which the blocks are configured and then wired together along the edges through the phosphodiester covalent attachment and through the center through stabilization provided bysteroid hydrogen binding.  During this encoding process, a separation event enables the separation of the encoded DNA structure into a double helix, and the steroid complex.  The steroid molecule may then bind onto the DNA strands, which will cause its separation.  The new access will be addressed by the encoded steroid structure to induce protein manufacturing through binding of amino acids.  After release of the protein, a new protein may be developed until the binding site of the steroid onto the DNA strands is disassociated. }
\end{figure}

An overview of the mechanistic flow is indicated in Figure \ref{fig:translation} to produce encoded proteins starting from basic tetra-ringed structures that are configured in a stacked format as in Figure \ref{fig:sterstack}.  Through a fabrication process, associating nucleotide pairing in conjunction with its paired steroid molecule, the phosphodiester connection is constructed along both edges of the structure.  At a certain point, the model becomes rigid, and the drive to the lower energy double helix induces a rotation of the formed nucleotides, ejecting the steroid and reducing the unified complex to just the double helix, which is more stable, but of much lower information content than the original unified complex.  

It is noted that the unified complex prior to its separation into a double helix and steroid molecules, would be a natural site for the accumulation of amino acids along the edge of the complex and down the center.  There is significant charge along the edges that would attract the carboxyl and amine groupings, and the spacing is appropriate as well, averaging roughly seven angstroms, which is consistent with the steroid to nucleotide spacing.  Thus, there would be significant building of the concentration of the amino acid surrounding the unified complex such that when separation occurs, assembly into amino acid chains can proceed immediately.

After the separation event occurs, the DNA is in a double helix format, but is accessible for transcription through the binding of the encoded steroid molecule onto the encoded nucleotide.  As indicated in Section \ref{sec:transcription}, the binding will open the strands for binding of the encoded steroid molecules onto half of the complementary base. There is an association of the amino acids to steroid molecules which is defined by the steroid molecules in conjunction with the double helix effectively defining a interaction vessel based upon size and chemical structure of the amino acid \cite{schaper2020design}.  This process induces the formation of encoded proteins as the fluidity of the steroidal attachment to the half base will enable the presentation of the amino acids within the interaction vessel, resulting ultimately in the chain propagation of the formation of the peptide bond.  

\begin{figure}[!htb]
\centering
\includegraphics[width=.8\textwidth]{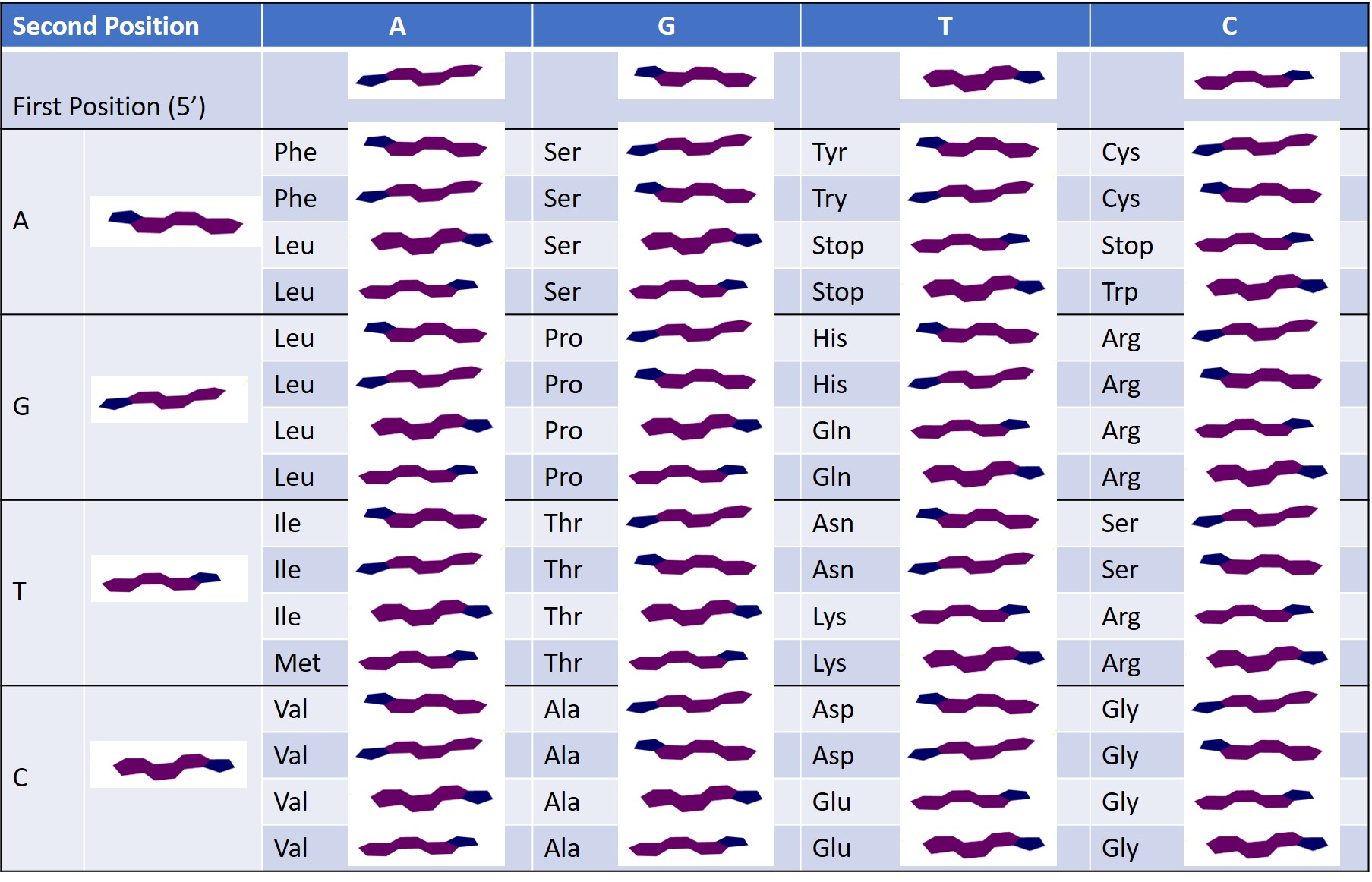}
\caption{\label{fig:stertable} The translation table for the orientation of the steroid molecules.  The translation table and the processes of forming the interaction vessel for the first DNA molecules is presented further in my recent book \cite{schaper2020design}.  }
\end{figure}

The structural association of the amino acid to the steroid molecule shows very precise matching.  The full set of 20 amino acid matches seems to be possible to assign the structural relation of the steroid structure, as it is a base four structure, to the amino acid structure to mirror that of the present-day code, although here the association is performed directly on the double helix.  As shown in  Figure \ref{fig:stertable}, it is possible to assign a structure for each of the amino acids.  As indicated in my book \cite{schaper2020design}, the match is based on the structural and chemical characteristics of the amino acid. 

When there is separation of the unified complex, the double helix will form, which requires a rotation of some of the nucleotides and an ejection of the steroid molecules and amino acids, connected through hydrogen bonding.  The availability of the DNA and the steroid molecules will come into association, and the strands will separate because of the stabilization force provided by the steroid molecule, which will focus the vibrational and oscillatory nodes of the strands to points surrounding the binding point.  The result is a separation of strands, and the binding of the steroid molecule onto the complementary bases of the nucleotides.  This will allow for the binding of the amino acids onto the steroid structure, and the result is an encoded protein structure, which can be replicated.  The double helix can also be repliated and error corrected using this procedure \cite{schaper2020design}.    

\subsection{Feedback Control}
\begin{figure}[!htb]
\centering
\includegraphics[width=.85\textwidth]{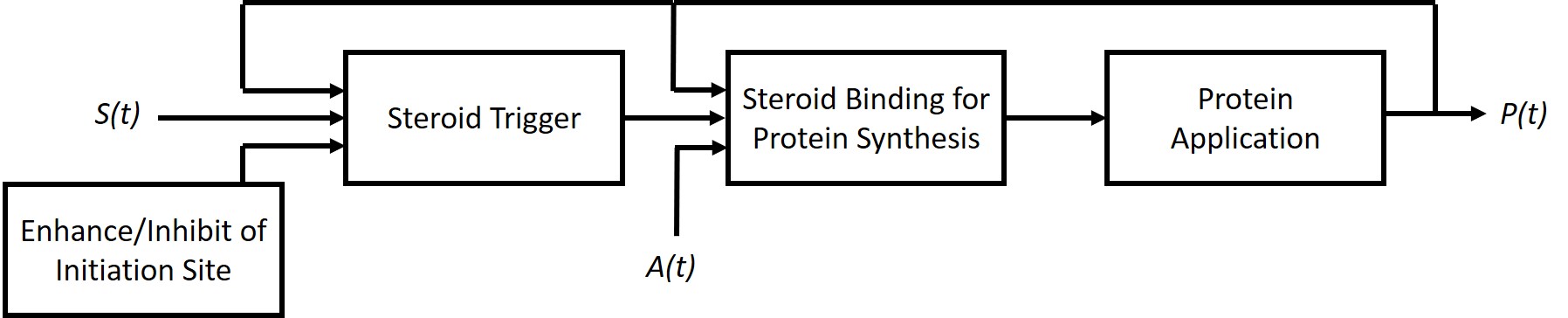}
\caption{\label{fig:feedback} The feedback loop to increase the rate of protein generation for those proteins that are more successful, that is find greater usage.  In this block diagram, the steroid molecule concentration triggers the strand separation, which is also a function of the enhancer and inhibitor concentration which may influence the trigger zone.  The protein molecule will also influence the steroid trigger, as a higher degree of use will place more force on the trigger point, and thus may disturb transcription.  In the figure, $S(t)$ denotes steroid structures capable of inducing transcription, $A(t)$ denotes amino acid concentration over time, and $P(t)$ denotes proteins that are synthesized and leave the DNA molecule for usage in the surrounding environment.}
\end{figure}

In addition to providing the mechanisms for generating proteins from amino acids, the steroid molecule, as an intermediary, has to have an ability to participate in a feedback loop to promote those amino acid sequences which are more likely to be of use in the surrounding environment, and thus induce increases in the availability of the molecule to further participate in molecular interactions.  In Figure \ref{fig:feedback}, the feedback loop is presented, in which the steroid molecule binding onto the strands induces strand separation, thereby accommodating transcription of amino acids into proteins.  If the proteins are utilized, it will exit the strand, and made the steroid structures available for further binding.  This feedback will continue onto the molecular structures inducing separation as well, as the forces from repeated implementation of the amino acid groupings on the strands may cause the degradation of the steroid strand separation element, and thereby terminate the strand separation and transcription.  Therefore, proteins that feedback to maintain or inhibit strand separation will also have an impact on the transcription rates.

As mediated through the steroid molecule, which functions for prebiotic applications to initiate transcription and as an intermediary for translation to enable protein generation, its regulation becomes a point at which cellular and later systems integration can take place.  Control of the concentration and transportation of steroid molecules into the cytoplasm and nucleus thus is able to differentiate the cellular objective, and therefore is itself a building block for further development ultimately resultant in systems performance.  The transition to tRNA from steroid molecules would provide a more structured environment for protein generation through separation of function.

\section{Discussion}
As a valid theory describing  the origin of DNA needs to be able to provide detailed answers to questions on its structure, the approach developed in this research article provides such answers both with regard to DNA nucleotides as well as steroid hormones, and even the development of proteins comprised of amino acids.    For example, questions that can be addressed are those that pertain to the  most basic of which as to how the base four code is fundamentally defined, extending to the most detailed, such as a quantifiable metric associated with the arc distance between nucleotide pairs, which correlates with the pitch of the stacked basis molecules, both approximately seven angstroms.  Moreover, there is agreement on the structural configuration, such as the rotation required to orient the nucleotides in correspondence to the present-day the double helix, which thereby enables a driving force to release the co-synthesized steroid molecules.  In addition, there is agreement in the combination of a ketone and a hydroxyl group to the phosphor element, which was needed in order to enable hydrogen bonding with both ends of the steroid molecules, as the group is typically expressed as PO$_4^-$.  Remarkably and simply, there were four orientations of one pair of steroid molecules to a second pair of orientations, thereby defining a base 4 code to DNA when formed as a sequential stack.  There is a match in the structural biology of the steroid molecule, steroid hormones, and paired nucleotide molecules:  It has been shown that the match is perfect and undeniable, formed by a trace of  precisely seventeen carbon, oxygen and nitrogen elements of nucleotide pairs.   And from this structural match, the mechanisms for the origin of DNA nucleotides and its related type of steroid molecules has been derived.  


Early in the process, the ``code of life" is defined, whereas alternative theories on the origin of life struggle with this concept, and here it is immediately defined.  This is a key step and insight, in which there is co-synthesis intermolecular development of the target and trigger.  Moreover, the parallel development of the DNA as target and the steroid molecule, and their simultaneous release and tied development, including the rotation of the nucleotides with the formation of the double helix as a driving force, constitutes a ``big bang" moment in the creation of life function.  As it resolves the ``chicken and egg" problem as to which one came first, whereby the answer is both and it was always tied since both were developed in an optimal intermolecular manner, is a satisfying answer.  By having the form of the trigger and target derive from the same structure, and intertwine in the simultaneous development, an optimized structure with respect to both molecular configurations is achieved.  

In developing this research, a striking result is the simplicity of the materials and in particular the process mechanisms involved in establishing the code of DNA nucleotide sequences.  Considering that DNA  developed soon after the formation of Earth according to present estimates, it had to be this way, that is simplistic and robust.   There is not the need for a lengthy evolutionary process requiring multiple building blocks to synthesize and put into practice, as the materials and functionality of DNA are correlated.  A confluence of unrelated probabilistic events is not required to generate molecular structures for lifeforms, or a sequential period of development of critical components: the development of the fundamental form of molecular life structure and function is achievable in a parallel yet linear fashion.  Moreover, the development of the steroid molecules at the same time as the DNA nucleotides also accelerates and simplifies the development of a critical life function.

To produce the coded structure and alternative outcomes, the methodology indicated in this article, in which two cyclic structures are paired, and then paired again, can be utilized to form other structures that can contain encoded larger molecules as well as their molecular function enablers.  This would be applicable to the formation of molecular storage devices, and the approach described in the article can be used to create alternative structures by manipulating the orientation of the molecules, and number of rings.  Furthermore, the approach described in this paper can synthesize at the same time both the storage medium, as well as the access mechanism through the parallel development of the steroid triggering molecule.  It is possible for precise replication of certain portions of the molecule through the use of the steroid triggering molecule, that is equivalent to transcription.

This methodology also should enable a questioning on how gene transcription may be performed.  The new approach to transcription involves the direct connection with the steroid molecule to bind onto the DNA nucleotide, as it is pre-programmed since both were developed in parallel, and one influenced the other through intermolecular binding through co-structural development.  Current strategy for transcription is through a receptor, hence indirect, but these results indicate that the steroid molecule is integrated with the DNA molecule, and is thus capable of inducing replication and transcription without the need for a receptor.  These ideas that steroids and the DNA nucleotides share the same structure, and thus there is likelihood that the binding of the steroid hormone to DNA nucleotides triggers gene transcription as first proposed and developed in the preprint \cite{schaper2020endogenous}, and these results on the origin of DNA should solidify those ideas.  The driving force for transcription is the stabilization energy provided by the steroid molecule that can be dissipated through destabilization about the bound area via lengthening of nearby hydrogen bonds along the nucleic acid strands.  Thus, if the current methodology of protein induced transcription is correct, then there was a early life mechanism of transcription that did not require proteins, but rather utilized steroidal molecules.

The results indicate that the reason for purine and pyrimidine as the bases that comprise DNA as follows: Purine and pyrimidine together account for three of the four rings of the steroid molecule, and the coordination through hydrogen bonds accounts for the fourth ring.    Yet another mystery that is further resolved by this work is the location of the missing hydrogen bond of the adenine-thymine pair relative to the cytosine-guanine pair.  While A-T has two internal hydrogen bonds, there is an available oxygen element on thymine that can form an intermolecular hydrogen bond with the mid-molecule hydroxyl group of corticosteroids, such as the steroid hormone cortisol.  Both results are further described in the preprints \cite{schaper2020endogenous, schaper2020structural}.  This result, which is verified experimentally since prednisolone possesses activity whereas prednisone has no activity prior to conversion to prednisolone, provides hard evidence that the symmetry between steroid hormones and DNA nucleotides is more than coincidental, but is in fact functional.  This result is important, as previously the A-T pairs were considered as ``weak", but the complex can be made strong when coupled with an intermolecular hydrogen bond from corticosteroids.

A major advantage of this approach is that the translation of proteins from amino acids is immediately defined and consistent with the overall framework.  Thus, the entire sequence of replication, transcription and translation is apparent, which is far further than alternative theories on the origins of DNA.  The areas where additional development would include some of the details associated with the specificity of the amino acid to the functional sites of the steroid molecule.  

The steroid molecule apparently is an inherent structure of DNA, as it is difficult to argue against that there is an embedded structure traced into each and every nucleotide pair comprising nucleic acids.   Thus, these results have  defined the structural basis of DNA as the steroid molecule.  There are a variety of similar structures within that family of molecules that can be used as the trigger for DNA to enable transcription.  Thus, steroid molecules should undergo close examination for health benefits than they currently are.  Corticosteroids are a first line of defense in inflammatory diseases, but other forms of steroids, including hormone replacement therapies, and other forms of synthetic steroid hormones should be closely studied, as currently there seems to be a barrier attached to these molecules.  Life function is inherently derived from the steroid molecule, and thus its use for therapeutic purposes should be closely assessed.   


The structural symmetry of steroid hormones and DNA nucleotides is a practical result, for example in the area of design of therapeutics.  It was indicated that each functional element of the steroid hormone can be assigned a purpose, such as the mid-molecule OH group to form an intermolecular hydrogen bond with thymine, the end groups to form an ionic coupling with the phosphodiester backbone, and the internal structures, such as rings, methyl groups, and double bonds to enhance stability and interaction.  Furthermore, the backside elements of the steroid hormones, such as an OH group, can influence the transcription process.  Indeed,  these concepts have been applied to RNA based applications in forming an intermolecular bond with uracil in a like manner of thymine \cite{schaper2020intermolecular}.

As the interaction of binding the steroid molecule onto the strands of the DNA to initiate strand separation, which would thereby enable replication and transcription, a type of molecular machine was synthesized.  It can be envisioned  that the DNA strands are in fluidic motion, and the pinching induced by the steroid elements, enabled a bubble to be generated on either side of the interactive zone, which would permit access by chemical replicating agents.  Depending upon the type of environment, the steroid triggering agents can be evolved to permit directed or sustained replication periods.  The transition to ionic bonding interaction of the phosphodiester chain of the steroid triggering molecule is an example of the need for sustained interaction with the DNA strands.  It is noted that the switch to PO$^-_4$ from PO$_4$H came after the release of the triggering molecules because it is not possible to associate two ions so close to each other without repulsion.

In developing a research paper like this, it is appropriate to speculate on the issue of evolution.    The relative ease whereby the DNA sequence is built, merely by stacking paired steroid structures, and that the steroid molecules as genetic replication and transcription, which were enabled by simultaneous released to form the double helix thereby establish a beginning for life function, the probability of producing multiple starting sequences consistent with different evolutionary pathways can be reasoned.  Further, while the results were developed such that the starting molecule was in the form of DNA, which permitted direct comparison, there is a potential likelihood that the first originating molecule of DNA did not contain a multitude of nitrogenous groups.  This more robust molecule would then take on a master role from which images of it are developed to produce the nitrogenous structure of DNA as known today. Moreover, the conditions required for producing the structures, namely the polycyclic aromatic rings and hydrogen bonding, are ubiquitous, and thus it is easy to imagine that such structures could be replicated in other environments besides that of Earth, and thereby deposited on Earth, which would be consistent with the discoveries of polycyclic aromatic hydrocarbons reported by the astrobiology community including chrysene\cite{colangeli1992raman}.  Thus, the process sequence results describing the origin of DNA as presented in this article are not inconsistent with many of the philosophical and scientific tenets.   

As presented by the checklist in the introductory section regarding the requirements for a valid new theory on the creation of life, the theory of the origin of DNA developed in this article satisfies each of the requirements.  It is a truly remarkable that  DNA nucleotides and steroid molecules were developed as a unified complex, developed in an intermolecular manner with each molecule influencing one another to enable a big bang, instantaneous  moment of the simultaneous synthesis of structure and function.  Thus, as soon as the steroid molecules are ejected from the DNA-steroid complex concurrent with the DNA nucleotides rotating into a position consistent with the double helix, a central life function results.

\section{Methods}
\begin{itemize}
\item Molecular modeling software:  The software program Avogadro was utilized to build the three dimensional molecular models.    The Avogadro software program was used to position the elements and calculate the bond distances.  An optimization routine based upon steepest descent was implemented to position the molecules automatically.  Matlab was used to generate the energy dissipation requirements when the DNA nucleotides were connected by the steroid molecule.   
\item Development of molecular models:  The DNA nucleotides and steroid molecules were developed and synthesized cooperatively in an intermolecular manner by stacking ten pairs which would result in TAGTC, which was selected to contain each letter and to arrange like  and unlike molecules next to each other with respect to purines and pyrimidines.  The intermolecular hydrogen bonds were configured, and then the optimization software was used to orient the molecules, positioning to minimize energy.  After establishing the stack, the phosphodiester backbone was assembled starting from the bottom at molecule 1, and then linking 3, 5, 7 and 9 by forming the N-C-O coupling, which will ultimately correspond to the nucleotide to sugar link.  After coupling, the sugar molecule was formed.  Both sides were completed, with hydrogen bonding stabilizing the steroid triggering molecule to the phosphodiester structure.  The internal nitrogen elements of the purine and pyrimidine components of the nucleotide pairs were then positioned accordingly, and the bonds were broken with the addition of oxygen groups selected at the appropriate position on thymine, cytosine, and guanine.  The methyl group was added to thymine.  The triggering steroid molecule associated with cytosine was designed to have the appropriate triggering molecule. Other developments of the steroid triggering molecule were deemed outside the scope of this study.  The rotation was performed on the nucleotide pairs that had a pyrimidine molecule on the  5' phosphodiester side for this example, so as to eject the co-synthesized steroid molecules.  Transcriptional positioning was performed by inserting the resultant steroid into the appropriate slot.  For the ionic study, two Mg$^{2+}$ ions were also included.  The optimization software was used to position the steroid molecule.     
\item Experimental: To obtain the relative activity of the synthetic steroids and steroid hormones, the literature was surveyed and representative values were selected for the activity of prednisone, prednisolone, cortisol, and dexamethasone \cite{labrin2011what}.  The experimental structures of these steroid hormones were then overlaid to the experimental structures of the A-T nucleotide pair to achieve a third hydrogen bond by intermolecular coupling.
\item Overall Procedure (journey):  The development of this theory took the path of a top-down investigation.  The initiating point was a mathematical analysis of a coordination of the nervous and cardiovascular systems to characterize a fever induced by pneumococcal  vaccination in conjunction with exercise \cite{schaper2019dynamics}.  An investigation into the chemistry of the interaction resultant in the fever led to the concept of inhibition of cortisol by prostaglandins, which also led to the concept of coupling two Ca$^{2+}$ ions at the glucocorticoid receptor to explain the differences in activity between the two molecules \cite{schaper2019competitive}.  In further pursuit of a fundamental understanding of a fever resulted in transporting the glucocorticoid receptor to the nucleus to evaluate possible transcription of cortisol and prostaglandins through ionic binding directly on the DNA nucleotides.  An evaluation of the structural symmetry of cortisol and the adenine-thymine base pair was determined, noting the intermolecular binding.  Thus the symmetry of steroid molecules and DNA nucleotides was discovered \cite{schaper2020endogenous, schaper2020structural, schaper2020new}.  This structural symmetry of steroid molecules and DNA nucleotides then led to the current results to define a basis for the transcription results perceived of cortisol on adenine-thymine, and other potential interactions between RNA nucleotide and other functional cyclic compounds.  Hence, the original mathematical basis describing the systems and cellular responses are consistent with the nuclear developments, and thus enable an integrated theory.  
\end{itemize}

\clearpage
\newpage

\begin{appendices}

\section{Reaction Mechanisms}
Reaction mechanisms are indicated to form the differentiated molecular structures of: (a) adenine-thymine; (b) guanine-cytosine; (c) steroid type paired with A-T; (d) steroid type paired with G-C.  The pairing of steroid molecules and DNA nucleotides works both ways, which means that the orientation of the steroid molecule impacts the synthesis of the DNA nucleotides, and vice versa, since the interaction of two results in blocking elements which prevent amination or reduction from occurring.  This section is presented with an indication of the synthesis steps that result in differentiating function, including the separation into the base pairs.  Then the suggested reaction steps for producing the final products is indicated, which includes the reduction of the steroid molecules.  Finally, the common steps are described, including the possible starting molecules from which the steroid molecules results.  There are other synthesis pathways than the ones suggested in this section, however, the overall logical sequence would be similar in terms of the transformation of the molecules into the final products.  It is important to note that the synthesis steps occur within the same environment, such that while one base pairing is being synthesized, another aspect may be taking place in parallel, and thus the overall reactive environment needs to be consistent across most steps.  At prebiotic conditions and environments, UV illumination will play a significant role and catalyze reactions for amination to take place at appreciable rates via free radical, photochemical reactions \cite{hoffmann2008photochemical}, as implied in the developments of this section.  In addition, as noted in Section \ref{sec:enzyme}, the reactions take place within three dimensional constructs, although some mechanisms are presented in a standard two dimensional framework.

\subsection{Enzymatic Coupling} \label{sec:enzyme}
As noted in reference \cite{filee2000origin}, in the original work of Watson and Crick speculated that the double helix of nucleic acids may have an enzymatic nature; but that idea did not take hold and since then the research direction has trended toward the linear logic of a synthesis starting from single-strand RNA nucleotides, which would ultimately lead to the DNA double helix complex.  However, based upon the analysis presented in this research article, it is possible to re-evaluate the speculation of Watson and Crick within an analytic framework of the origination of DNA as an enzymatic construct of interleaved tetra-ringed structures, of which one section is covalently bonded to a phosphodiester backbone, while the adjacent structure is connected through hydrogen bonding in close proximity.

As indicated in Figure \ref{fig:enzyme}, the coordination of the paired building blocks enables a sequence of effective reaction vessels, secured at the ends through phosphodiester covalent coupling, and through the center by hydrogen bonding to amino acids.   This confined environment enables the potential for an enzymatic capability in which the temporary formation of bonding arrangements to permit intermolecular cyclic structures that can establish element substitution, the degradation of chemical bonds to reduce intramolecular stress, and the  differentiation of the templates into their individual capacity as a nucleotide base.  While the structures did have aromatic stabilization, the intermolecular structures may also have benefitted from resonance stabilization, and thus the intermolecular constraints to maintain an in-plane structure during the synthesis steps would have promoted the present-day configuration in which the original aromatic structure was traded for an amine based self-assembled construct of a planar configuration.


\subsection{Linking to the phosphodiester}
\begin{figure}[!htb]
\centering
\subfloat[Amination]{\includegraphics[width=.18\textwidth]{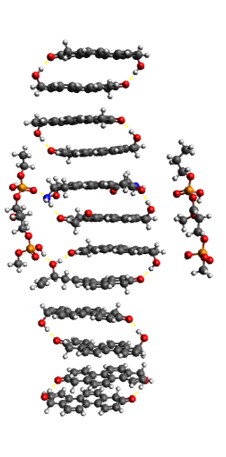}}
\subfloat[Connect]{\includegraphics[width=.2\textwidth]{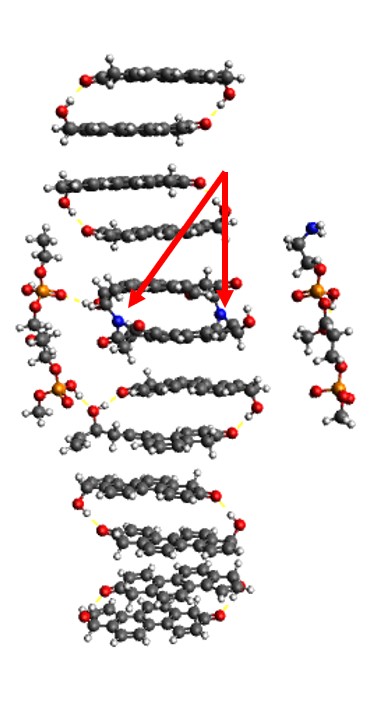}}
\subfloat[Separate]{\includegraphics[width=.2\textwidth]{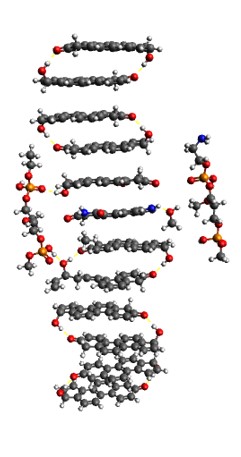}}
\subfloat[Bond]{\includegraphics[width=.17\textwidth]{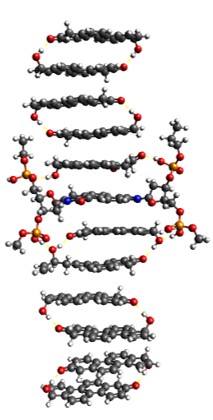}}
\caption{\label{fig:couple} (a)  To initiate the link of the phosphodiester backbone to the DNA nucleotides, the two amine groups are attached to the molecule, which will become the steroid molecule. (b) a binding is achieved to replace the alpha carbon between the diketone groups, and on the other end, the carbon binding to the hydroxyl group is replaced. (c) after binding, there is tendency for leveling with respect to the steroid molecules above and below the molecule where the amine group was replaced.  (d) the final orientation, and ready for further amination and separation.}
\end{figure}

In Figure \ref{fig:couple}, the method of linking stacks of molecules together is indicated, in which the enzymatic characteristics of the implicit reaction vessel are implemented.  The initiating step involves the amination of what will be the steroid molecules, that is the non-nucleotide molecule.  This occurs at both ends of the molecule, in proximity to the diketone group on one side and the hydroxyl group on the other side, as indicated in Figure \ref{fig:couple}(a).  Because this planar structure will enable a catalytic activity, the amine group reacts with both of the ketone groups, and the amine group achieves alkene addition on either side of the phenol group, as shown in Figure \ref{fig:couple}(b), which then integrates within the DNA nucleotide molecule, displacing the residual as methane and as methanol.  This reaction is simliar to the azofullerene synthesis \cite{reuther2000synthesis,vostrowsky2006heterofullerenes,li2016fullerene}.  The connection is then achieved by binding onto the inserted amine groups, and then the formation of the sugar group to stabilize the phosphodiester backbone, as indicated in Figures \ref{fig:couple}(c) and \ref{fig:couple}(d), respectively.

\subsection{Differentiation of A-T and G-C Templates}
The implementation of this configuration involves intermolecular hydrogen bonding to cross-couple the two pairs, which  is critical in defining the differentiation between A-T and C-G pairing.  Since the ends are already coupled to connect the pairs, hydrogen bonding pairs are connected through the middle of the molecule.  The ketone group is configured two carbons away from the ketone group associated with the end-based pairing.  This ketone group is then connected to a hydroxyl group connected in proximity to the ketone group on the other molecule.  The position of the ketone group relative to the adjacent steroid molecule of the adjacent pair will define whether the molecule becomes A-T or G-C.   Interestingly, the ketone group associated with the A-T pairing of the steroid molecule will cross-couple to the nucleotide molecule and reduce the chances of amination from occurring, and thus arrive at an adenine.  This is indicated in Figure \ref{fig:sterstack}(a) and (b) for G-C and A-T respectively, indicating the position.  This is demonstrated by looking at the electrostatic potential in Figure \ref{fig:sterstack}(c) and (d). This provides the underlying basis for the differences between A-T and G-C.

\begin{figure}[!htb]
\centering
\subfloat[G-C]{\includegraphics[width=.45\textwidth]{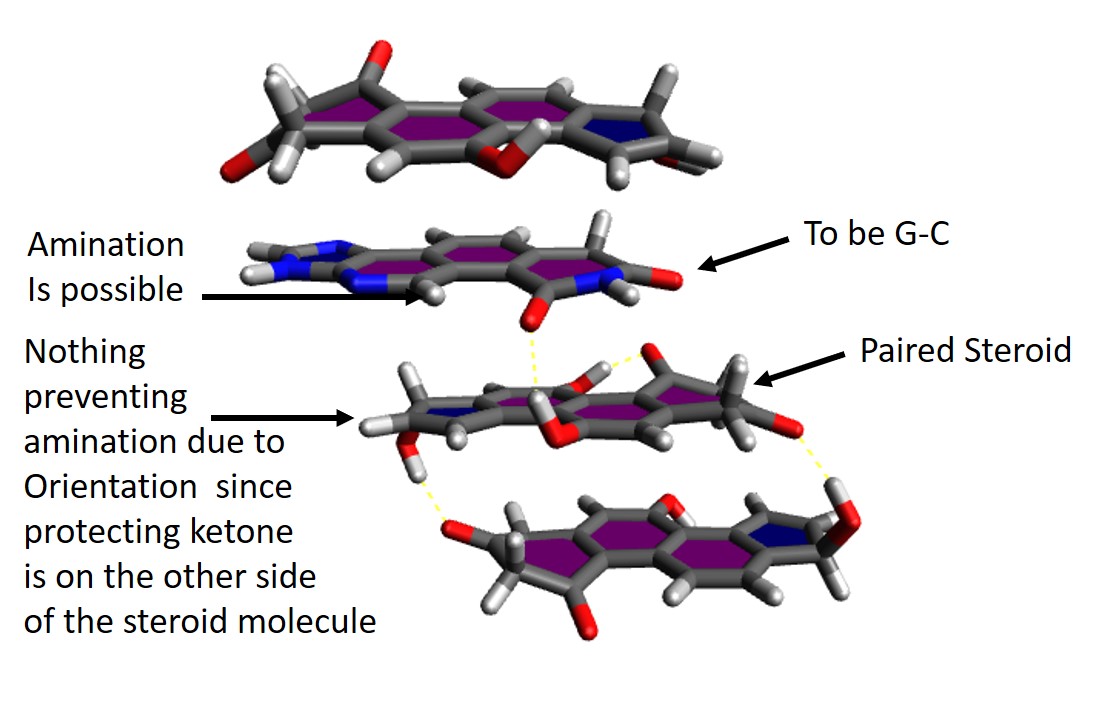}} \hfill
\subfloat[A-T]{\includegraphics[width=.45\textwidth]{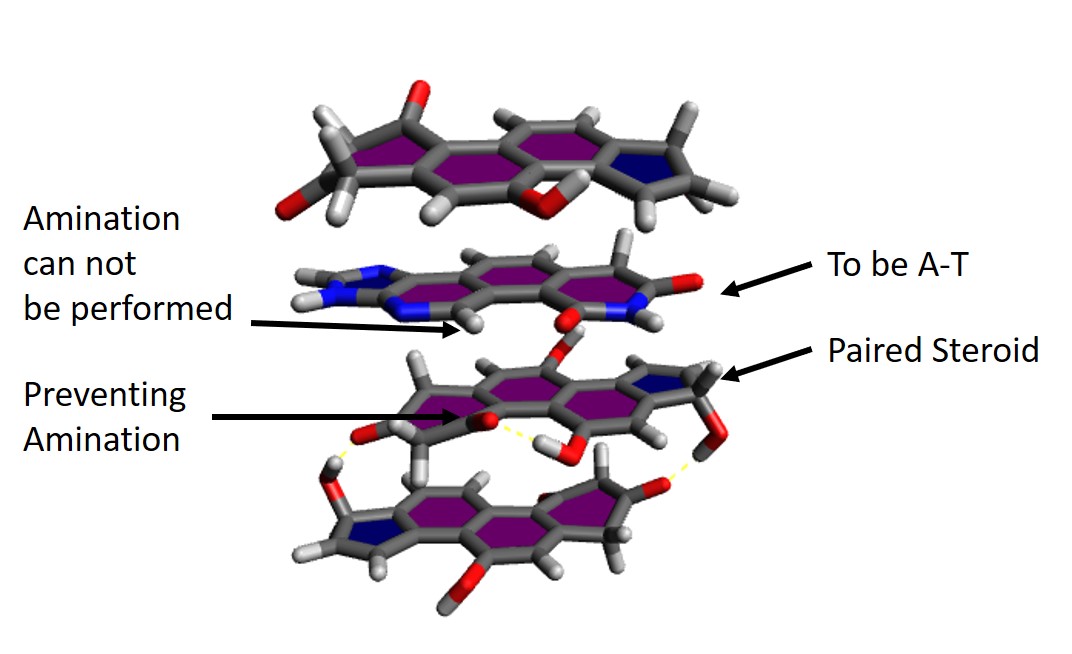}} \hfill \\
\subfloat[G-C]{\includegraphics[width=.45\textwidth]{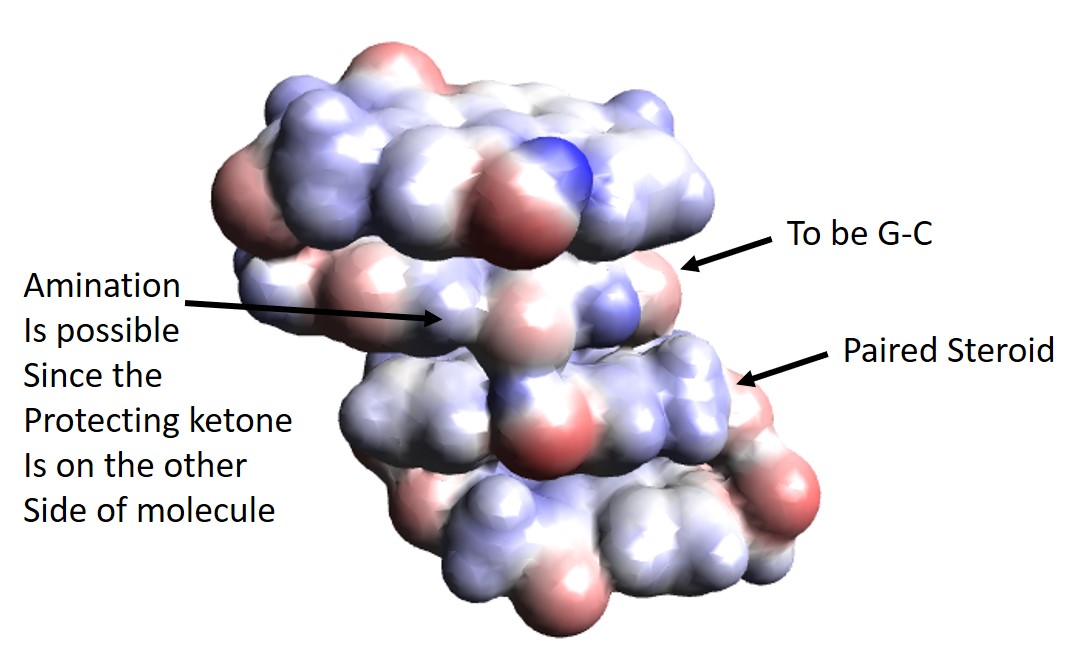}} \hfill
\subfloat[A-T]{\includegraphics[width=.45\textwidth]{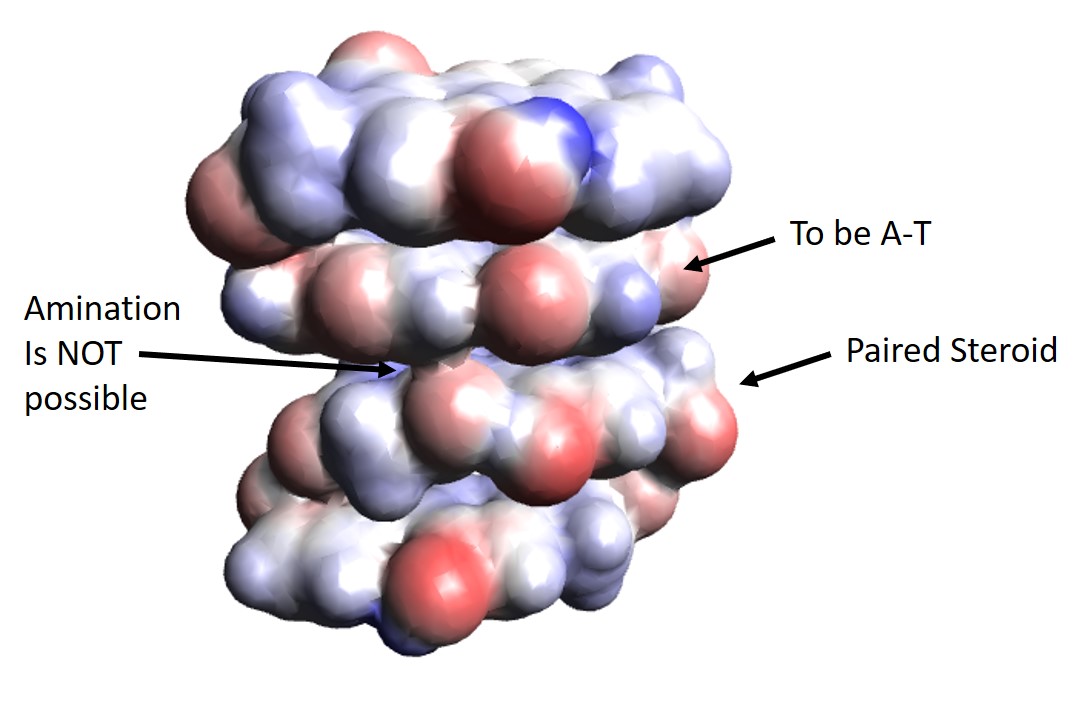}} \hfill
\caption{\label{fig:sterstack} (a) For G-C, the orientation of the steroid molecule, second from the bottom, with respect to the molecule that will become a G-C nucleotide is oriented such that amination is possible on the leading element of guanine, as initiated through UV illumination.  Thus, an NH$_2$ group will be affiliated with the leading side of guanine. (b) For A-T, however, there is a connecting group of O-OH associated with the steroid molecule pairing, thus amination is not possible.  Ultimately, this is why A-T has two internal hydrogen bonds, while G-C has three.  (c) The electrostatic potential indicates that the steroid molecule will not influence the G-C amination.  (d) The electrostatic potential indicates the coupling with the steroid molecule which will prevent amination for A-T.}
\end{figure}

\begin{figure}[!htb]
\centering
\subfloat[G-C]{\includegraphics[width=.45\textwidth]{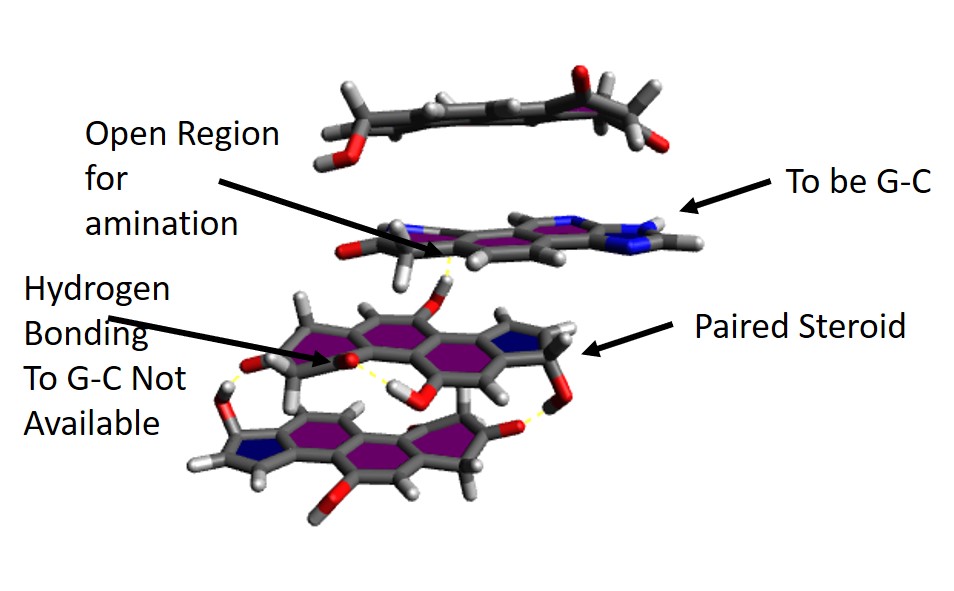}} \hfill
\subfloat[A-T]{\includegraphics[width=.45\textwidth]{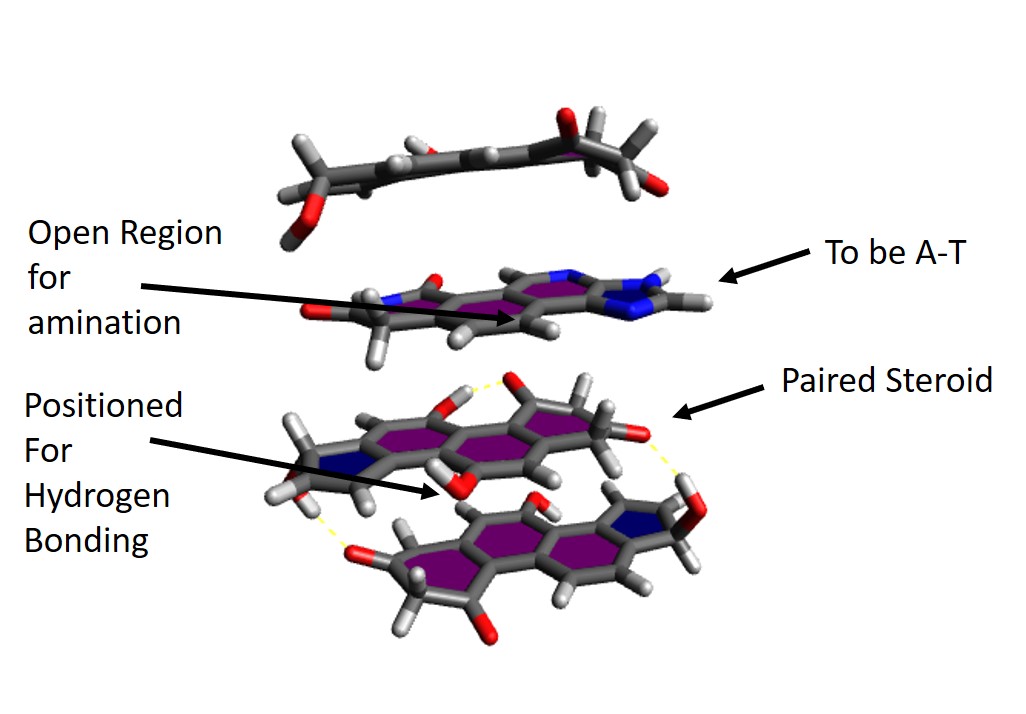}} \hfill \\
\subfloat[G-C]{\includegraphics[width=.45\textwidth]{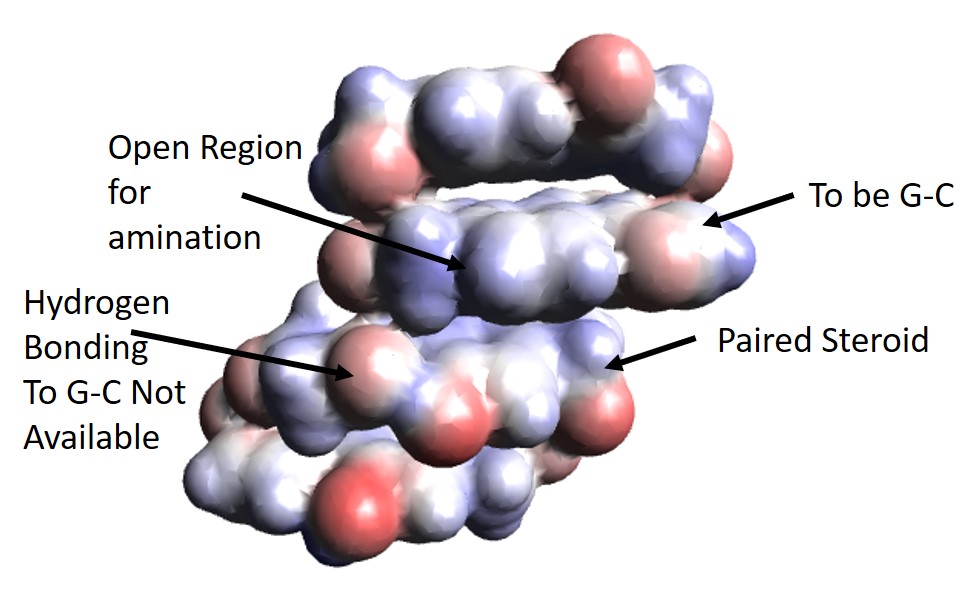}} \hfill
\subfloat[A-T]{\includegraphics[width=.45\textwidth]{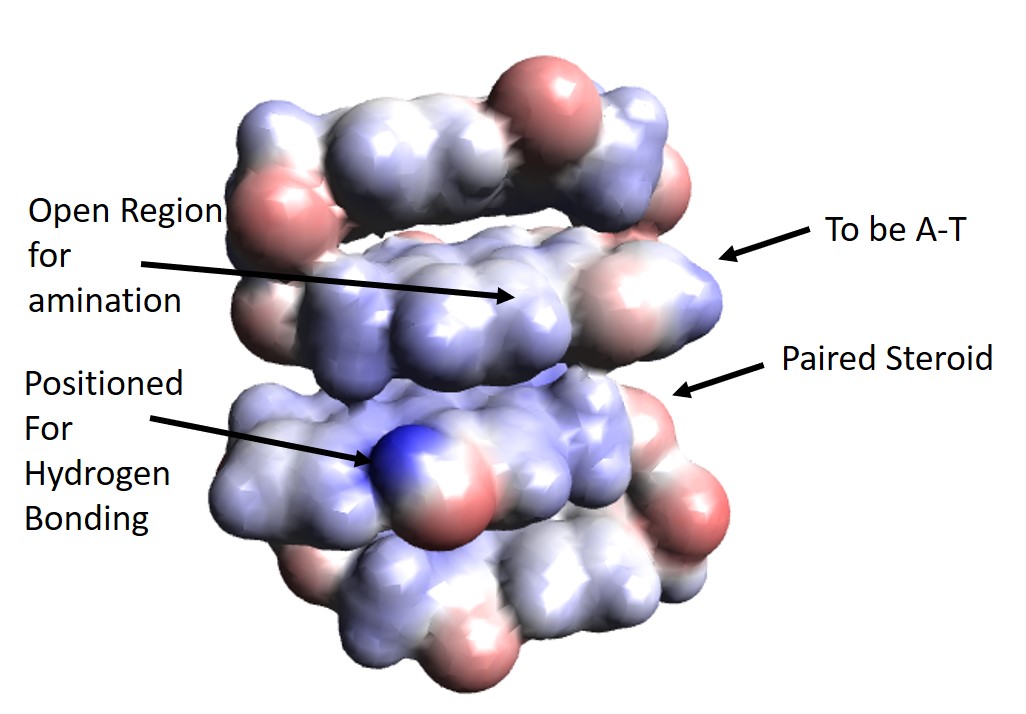}} \hfill
\caption{\label{fig:backstack} On the other side of G-C relative to Figure \ref{fig:sterstack}(a), the orientation of the steroid molecule, second from the bottom, with respect to the molecule that will become a G-C nucleotide indicates potential influence over the oxidation and amination. (b) For A-T, there is little association with the backside other than a potential hydrogen bond, and thus oxidation will be effective at the thymine molecule side.  (c) The electrostatic potential indicates that the steroid molecule will  influence the G-C association on the backside.  (d) The electrostatic potential indicates the coupling with the steroid molecule will influence oxidation for A-T.}
\end{figure}

In addition, this issue is also seen on the other side of the molecules, where the access of the molecules, and their relationships between the paired relationship of the steroid molecule to the nucleotide molecule, which will ultimately become the DNA nucleotide base pairing.  In Figure \ref{fig:backstack}(a) and (b), the other side of the steroid - DNA complex is indicated.  The electrostatic potential indicates the interaction of the steroid agents and the molecule that will become the DNA nucleotides in Figure \ref{fig:backstack}(c) and (d) for the G-C and A-T pairings.  In addition, the pairings  for C-G and A-T also show similar results, both on the front side and the back side.

\subsection{Procedure}

\begin{figure}[!htb]
\centering
\includegraphics[width=.4\textwidth]{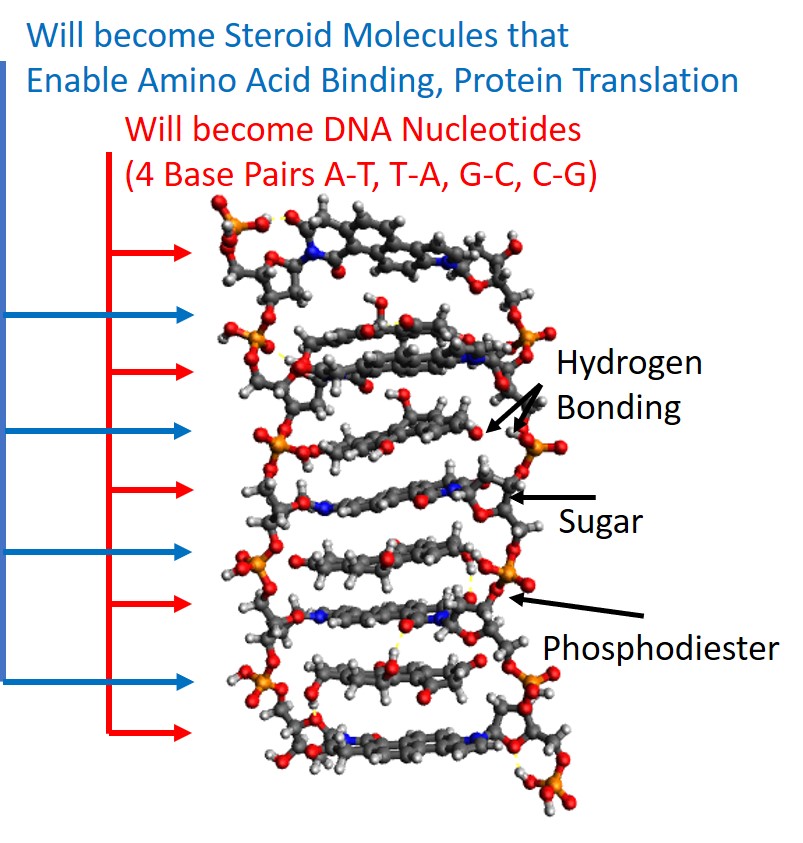}
\caption{\label{fig:DNAOH} Model indicates the paired synthesis of the steroid molecules and the DNA nucleotides during a stage of fabrication where the DNA nucleotides are formed, but not yet separated into their basic base pairing with associated hydrogen bonding.  There is interaction with the steroid molecule during the formative stages, and the DNA nucleotides influence the formation of the steroid molecule, and the steroid molecule influences the DNA nucleotides.  The association with amino acids is not included here. }
\end{figure}

In binding the phosphodiester backbone prior to its disassociation, as shown in Figure \ref{fig:DNAOH}, of which the binding occurs at both ends of the nucleotide molecules 1, 3, 5, 7 and 9, while hydrogen bonding stabilizes the steroid  molecules 2, 4, 6 and 8.  Note that phosphor group contains both OH and an O, such that hydrogen bonding can be achieved at both ends of the steroid molecule:  one side of the steroid molecule contains an OH group which will hydrogen bond with the O element of the phosphate group, and on the other side, the steroid molecules comprises a ketone group which will hydrogen bond to the hydroxyl group of the phosphate group.  The conversion of the link to a nitrogen element is indicated, and the binding occurs either at the middle carbon located between two ketone groups of the six-membered ring, which is discussed in the subsequent sections.    It is noted that there is sufficient space for occupation of the steroid structure in between each DNA nucleotide, as both structures are relatively planar.

Continued amination, oxidation, and reduction steps are performed to complete the co-synthesized fabrication of not only the encoded DNA nucleotides, but also the steroid molecules.  These steps are described in the following section.  Interestingly, at the end of the synthesis steps, the steroids are released from the unified complex when the pyrimidine molecules on the 5' side rotate into position to form the double helix.  Thus, there is a culminating step to the synthesis which brings both molecules to life, as the steroid molecules can interact with the DNA double helix to enable replication and transcription.

The overall procedure to produce a differentiated structure is indicated in Figure \ref{fig:enzyme} for the adenine-thymine pairing.
In Figure \ref{fig:enzyme}(a), the differentiation and separation of the DNA nucleotides is seeded by the amination onto the alpha hydroxy of the steroid molecule, with subsequent attack to the DNA nucleotide in a position to aminate the rings of the purine molecules.  Figure \ref{fig:enzyme}(b), an amination at the steroid and a coupling to the C ring of the DNA nucleotide, when still intact is achieved to initiate the separation. Figure \ref{fig:enzyme}(c) completion of the amination, and decoupling of the B ring with oxidation onto thymine completes the process.  Figure \ref{fig:enzyme}(d), with the removal of the ethene molecules, there is self alignment of the pairs to complete the transformation

\begin{figure}[!htb]
\centering
\subfloat[Initial]{\includegraphics[width=.2\textwidth]{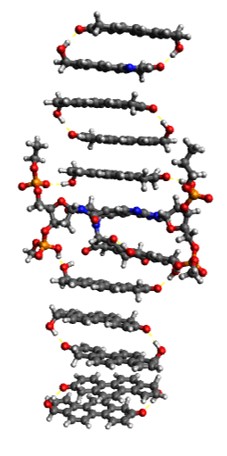}}
\subfloat[Bonding]{\includegraphics[width=.2\textwidth]{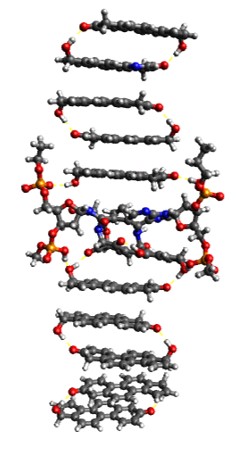}}
\subfloat[Aligning]{\includegraphics[width=.2\textwidth]{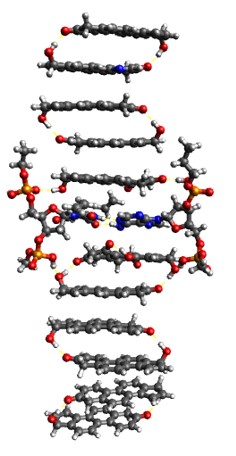}}
\subfloat[Final]{\includegraphics[width=.18\textwidth]{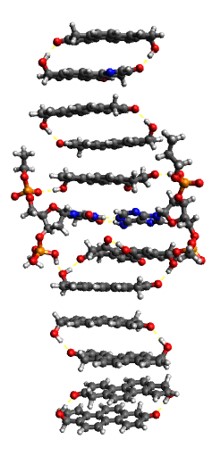}}
\caption{\label{fig:enzyme} (a) The differentation and separation of the DNA nucleotides is seeded by the amination onto the alpha hydroxy of the steroid molecule, with subsequent attack to the DNA nucleotide in a position to aminate the rings of the purine molecules.  (b) An amination at the steroid and a coupling to the C ring of the DNA nucleotide, when still intact is achieved to initiate the separation.  (c) Completion of the amination, and decoupling of the B ring with oxidation onto thymine completes the process.  (d) with the removal of the ethene molecules, there is self alignment of the pairs to complete the transformation. }
\end{figure}

\subsection{Separation into A-T and G-C Templates}

\begin{figure}[!htb]
\centering
\subfloat[Formation of A-T]{\includegraphics[width=.8\textwidth]{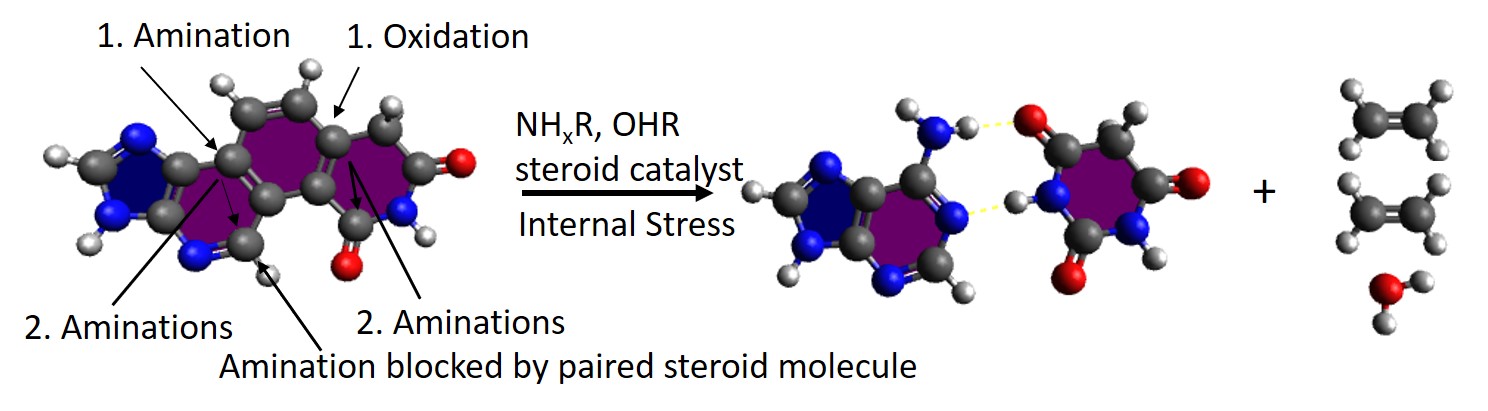}}\\
\subfloat[Formation of G-C]{\includegraphics[width=.8\textwidth]{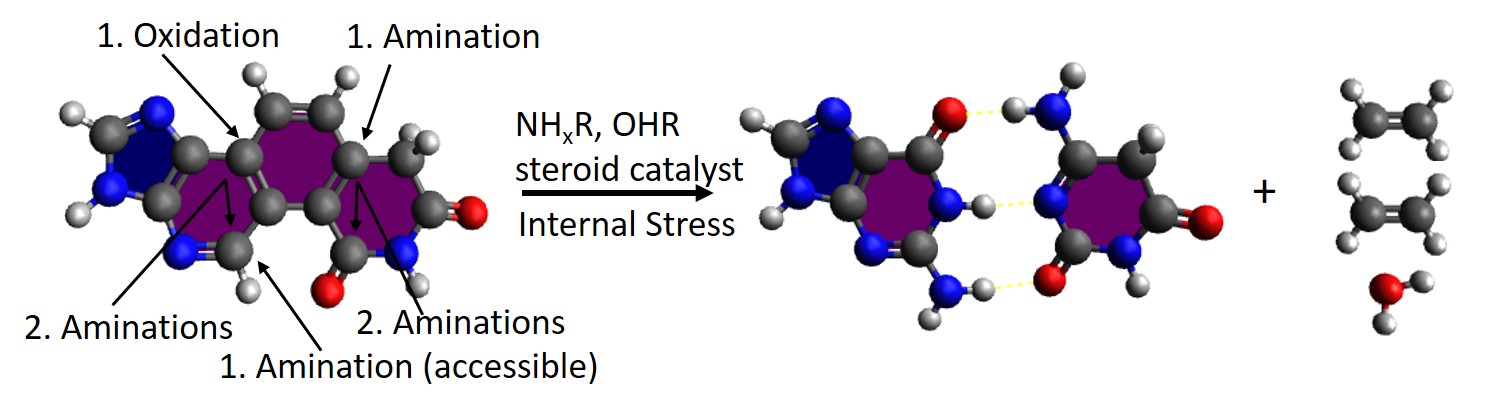}}
\caption{\label{fig:ATsynth} (a) For the A-T base formation, it is important to recall the result that indicated because of the paired steroid molecule positoning of its ketone elements, it is not possible the aminate the front of the molecule, and the amination occurs on the purine side, adenine.  This enables the oxidation at the thymine side.  The separation then occurs though an amination on both sides of the molecule, with the double bond established on adenine. (b) For the G-C base formation, it is also important to recall the result that indicated because of the paired steroid molecule positioning of its ketone elements, it is  possible to aminate the front of the molecule, and the amination occurs on the purine side, guanine.  This orientation also enables the oxidation at the guanine side, and hence amination at the cytosine side.  The separation then occurs though an amination on both sides of the molecule, with the double bond established on cytosine.}
\end{figure}

As indicated in Figure \ref{fig:ATsynth}(a), the reaction mechanism suggested for the formation of adenine-thymine is indicated.  It is important to note that this reaction takes place in conjunction with pairing to the steroid molecule, which has hydrogen bonding to the front portion of the adenine-thymine group, which prevents amination from occurring on the adenine front side.  This provides a key differentiating step.  In addition on the backside, the interaction leads to the preference for oxidation on the thymine side.  Aminations in an ammonia environment, with availability to oxidation components is performed.  The entire structure is under stress, in which there is a significant driving force to form the double helix, and thus enable the separation of the base pairs.  For the G-C base formation, the suggested reaction mechanism and strategy is indicated in Figure \ref{fig:ATsynth}(b), which occurs within the same environment as that of A-T.  Recall that because of the paired steroid molecule positioning of its ketone elements, it is possible to aminate the front of the molecule, and thus there is amination on guanine.  This enables the oxidation at the thymine side.  As with A-T, the separation is driven by the drive to form the double helix, and its lower energy state. 

Thus, this provides the basic answer as to why the A-T pairing only has two internal hydrogen bonds, while the G-C pairing has three internal bonds:  because of the relative orientation of it to the paired steroid molecule permits hydrogen bonding.  Moreover, it indicates why the overall structure of the base pairings occurs, and provides the driving force for it to occur, high stress, and a significant reduction in energy by transitioning to a double helix form.   It will also provide answers to why NH2$_2$ is on the pyrimidine purine side for G-C but on the purine side for A-T, and other such comparative matters.  It is important to note that these reactions are effectively taking place over the entire molecule, so the same process steps have to be addressed concurrently on how they would react to A-T if working on C-G, for example.

\subsection{Differentiated Final Products}
\begin{figure}[!htb]
\centering
\subfloat[Formation of Thymine]{\includegraphics[width=.8\textwidth]{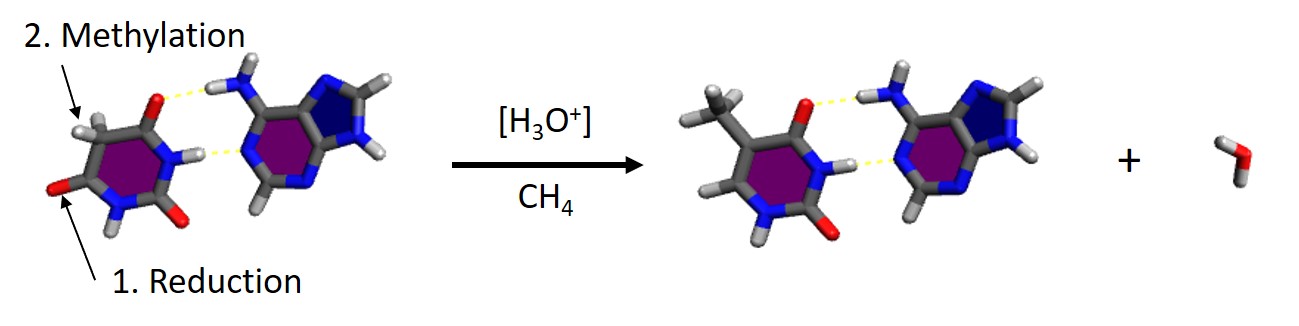}}\\
\subfloat[Formation of Cytosine]{\includegraphics[width=.8\textwidth]{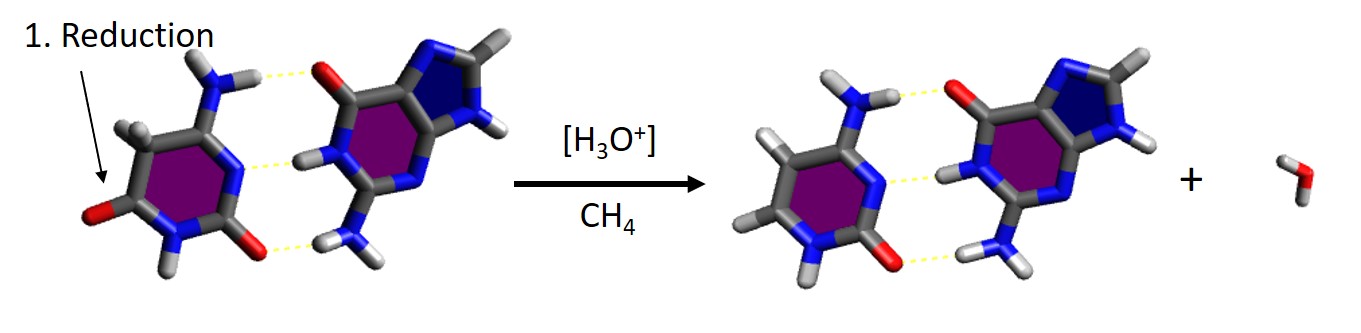}}
\caption{\label{fig:thymine} (a) Steps to reduce the oxygen group on thymine and the methylation to form the methyl group associated with thymine.  The oxygen group of thymine is not engaged in hydrogen and has an adjacent available carbon segment which will allow for its reduction. (b) The reaction sequence reducing the oxygen functional group of cytosine.  The positioning of the oxygen group will enable its reduction, whereas the other segments are engaged with the guanine base.}
\end{figure}

The final steps to produce the final products are all performed in the same reducing and methylation environment.  In Figure \ref{fig:thymine} the steps to reduce the oxygen group on thymine is indicated followed by the methylation to form the methyl group associated with thymine.  Since the oxygen group of thymine is not engaged in hydrogen bonding and has an adjacent available carbon segment, this will allow for its reduction. In Figure \ref{fig:thymine}(b) the reaction sequence is indicated to reduce the oxygen functional group of what will become cytosine.  The positioning of the oxygen group enable its reduction, whereas the other segments are engaged with the guanine base, and thus do not undergo reduction.

\begin{figure}[!htb]
\centering
\subfloat[Steroid for A-T pairing]{\includegraphics[width=.7\textwidth]{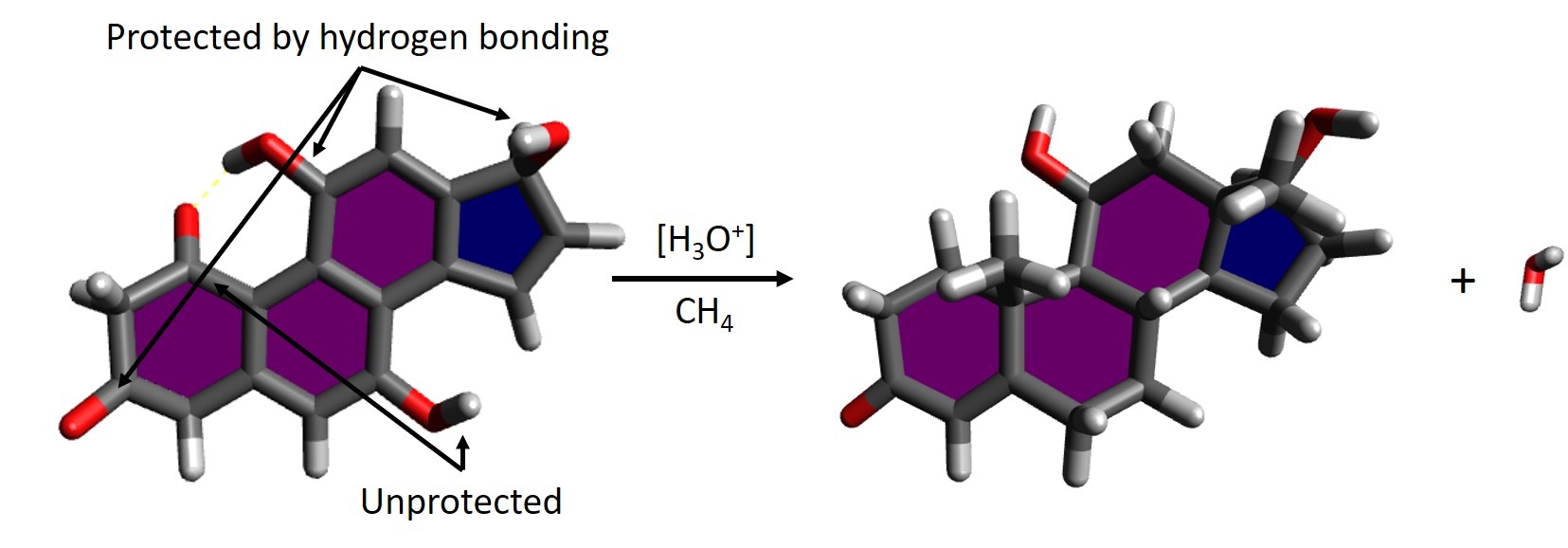}} \hfill \\
\subfloat[Steroid for G-C pairing]{\includegraphics[width=.7\textwidth]{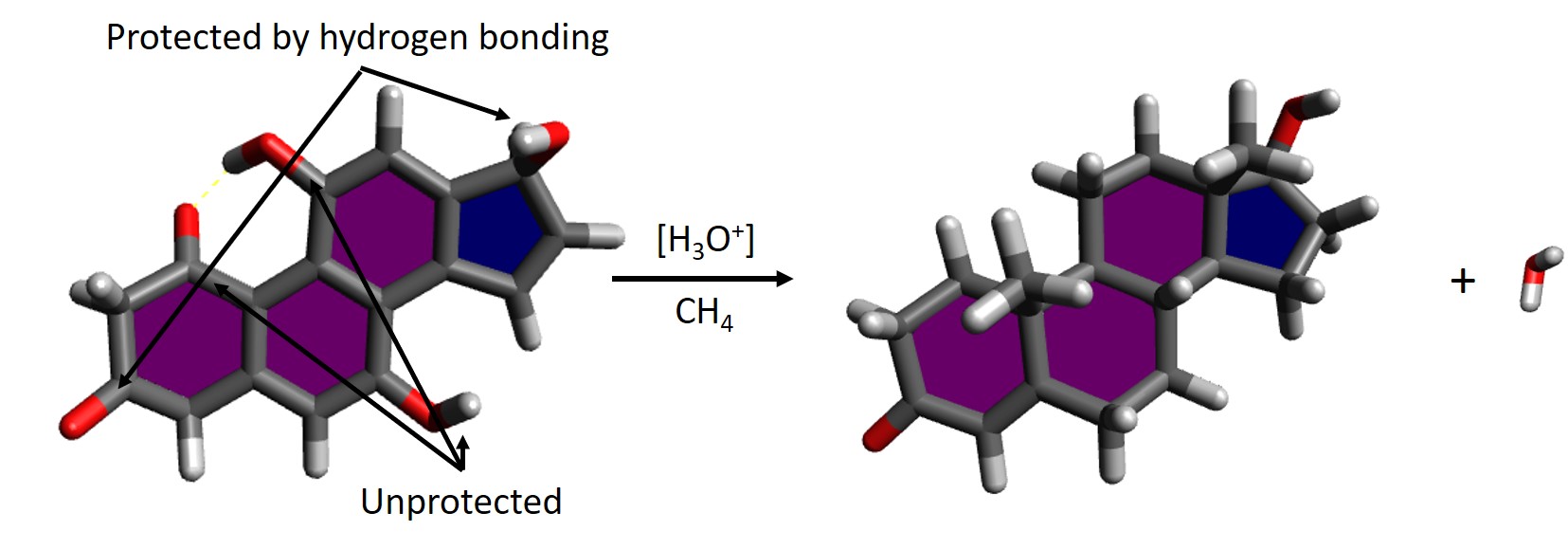}}
\caption{\label{fig:steroids} (a) For the steroid molecule paired with A-T, the reduction and methylation will take place on areas that are not protected by the interaction with A-T.  This will result in the reduction of double bonds and elimination of some of the oxygen functional groups.  The ends will remain intact since there is hydrogen bonding interaction with the phosphodiester backbone. (b) A similar result is obtained for the steroid molecule paired with C-G, except that because the mid-molecule oxygen groups are not engaged with C-G, they will be reduced.  Hence, there are two types of steroid molecules produced. }
\end{figure}

Finally, and in parallel, in Figure \ref{fig:steroids}(a) the steroid resultant in the corticosteroid function is indicated, as it will also undergo reduction and methylation, while the thymine and cytosine groups are reduced.  Because a portion of what will become the corticosteroid molecule is engaged in hydrogen bonding with A-T, it will survive the reduction process.  While in Figure \ref{fig:steroids}(b), those molecules associated in hydrogen bonding with C-G will be reduced, and thus not contain a oxygen group.    Hence, there are two types of steroid molecules produced: one associated with A-T and T-A, and the other associated with G-C and C-G.

\subsection{Starting Materials} \label{sec:ch}

\begin{figure}[!htb]
\centering
\subfloat[Conversion of chrysene]{\includegraphics[width=.7\textwidth]{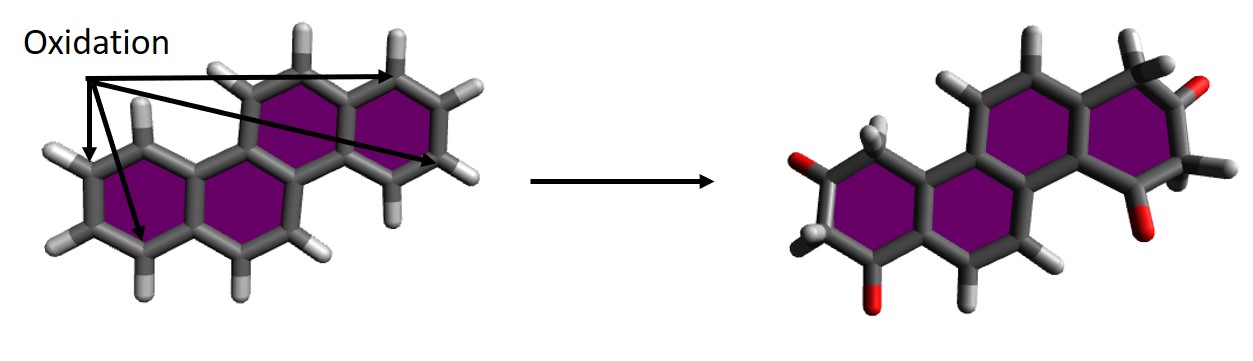}}\\
\subfloat[Steroid complex]{\includegraphics[width=.7\textwidth]{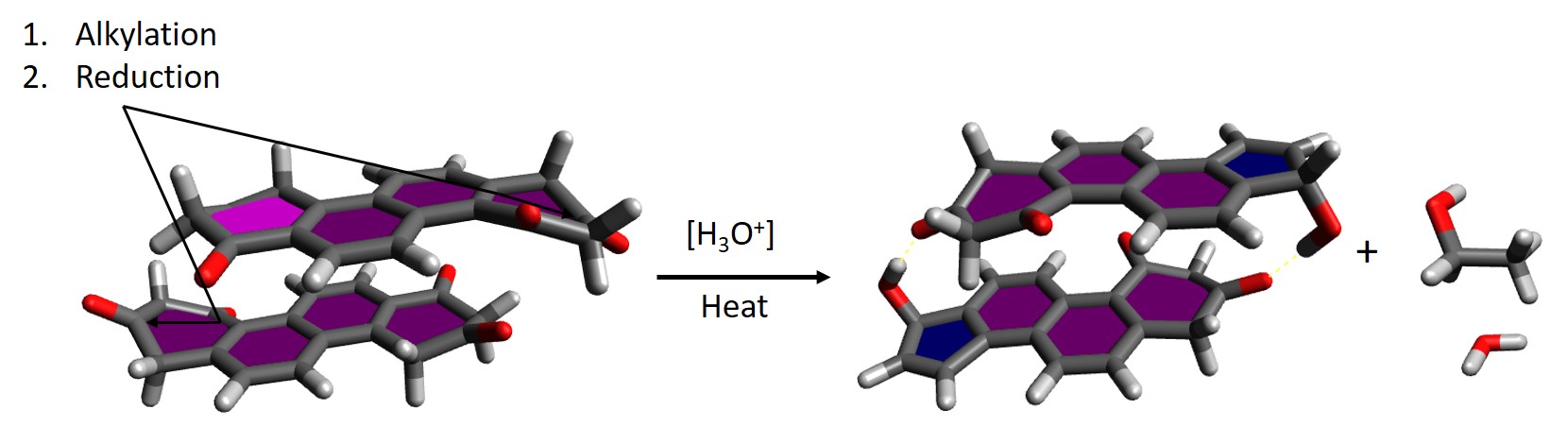}}
\caption{\label{fig:chyr}(a)  Conversion of chrysene to a diketone on both sides of the molecule.  In between the ketone groups are acidic hydrogens. (b) The reaction to reduce the chrysene based structure to the steroid structure is indicated.  }
\end{figure}

To broaden the scope, the developments will consider a related structured, the fused aromatic four ring structure of chrysene, indicated in Figure \ref{fig:chyr}(a), which is a relatively simple organic structure.  From a structural perspective, the molecule can be aligned with itself under several different configurations, rather than just one if it were to be linear.  It is also a stable compact structure, and further is the same general structure of a steroid molecule, is consistent with the general structure of both the DNA nucleotide pairs and the steroid hormones.  With the addition of the functional groups to chrysene, stacking of the molecules can be secured through attraction of the alpha acidic group situated in between the two ketone molecules.  Further, this molecule is active and can be reduced as indicated Figure \ref{fig:chyr}(b), which would be enabled through a catalyst, as with all the reactions, and includes photochemical effects. 

\subsection{Ionic Binding}
\begin{figure}[!htb]
\centering
\subfloat[Internal Hydrogen Bond]{\includegraphics[width=.3\textwidth]{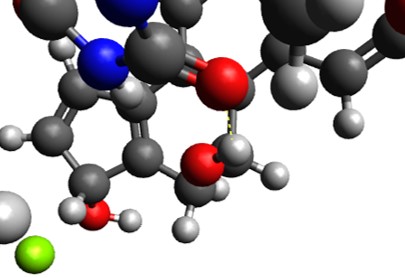}} \hfill
\subfloat[Electrostatic Potential]{\includegraphics[width=.2\textwidth]{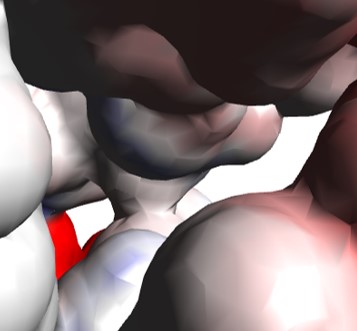}} \hfill
\subfloat[Ionic Binding (L)]{\includegraphics[width=.2\textwidth]{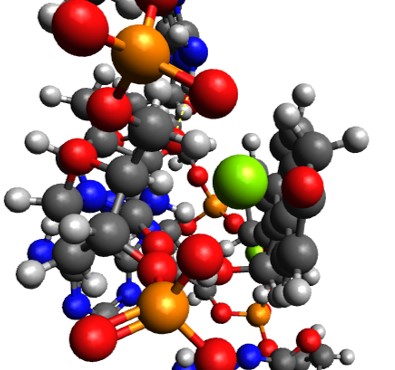}} \hfill
\subfloat[Ionic Binding (R)]{\includegraphics[width=.21\textwidth]{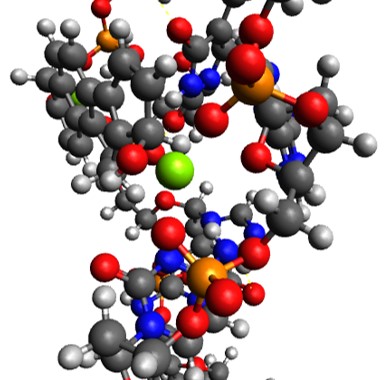}} \hfill
\caption{\label{fig:DNAbond} (a) Intermolecular hydrogen bonding distances are achieved by the  steroid to thymine coupling, as well as the internal thymine to adenine hydrogen bonding. (b)  The electrostatic potential of the intermolecular hydrogen bond between the steroid triggering molecule and thymine is indicated, along with the internal two hydrogen bonds of adenine to thymine. (c) Ionic coupling of the end group of the steroid molecule, which will actually be a steroid hormone is achieved.  (d) The two end groups of the steroid molecule are ionically coupled to the nucleotide pairs through adjacent phosphodiester groups.}
\end{figure}

As indicated earlier, the original DNA was of a phosphate that had an OH group instead of oxygen ion.  The hydrogen group was used for hydrogen bonding to the steroid molecule, which contained a ketone end group as well as a hydroxyl end group.  However, modern examinations of the DNA molecule assign the phosphate group as PO$^-_4$ an ionized representation.  Thus, apparently there is a slight shift of the DNA molecule from transcription through hydrogen bonding of the steroid element to the ionic bonding of the phosphate group, which is much stronger, and this structure will be used here to evaluate DNA transcription.  For the modern representation to indicate its usage with Mg$^{2+}$ ions, of which other ions can be utilized of a positive two charge to link with adjacent phosphate groups, and include Ca$^{2+}$, Mn$^{2+}$ and Zn$^{2+}$.   In Figure \ref{fig:DNAbond}(b), the intermolecular hydrogen bonding between the steroid molecule and the thymine-adenine pairing is indicated through examination of the electrostatic potential. The intermolecular hydrogen bonding between the steroid molecule  is indicated, which is 1.827 {\AA}. The ionic binding is tight, with the distance between the ion to the steroid group, for example the ketone group is 1.972 {\AA} and the oxygen element of the phosphodiester backbone is 1.699 {\AA} and 1.673 {\AA}, as indicated in \ref{fig:DNAbond}(c) and (d).   For the simulation results, the energy of the complex prior to binding was -1,161 kJ/mol, and after binding, it was -9,010 kJ/mol, thus providing for 7,849 kJ/mol of energy to dissipate through bond breaking.  

\clearpage
\newpage

\section{Conflicts of Interest}
No external funding was used for this research.  

\section{Provisional Patents} \label{app:patents}

The author has filed several personal provisional patents associated with this line of research.  These provisional patents include:
\begin{enumerate}
\item Charles Schaper, ``Binding Steroid Molecules to DNA", Provisional Patent: US 62/977,216. Filed 3/9/2020.

\item Charles Schaper, ``Binding Nucleic Acids", Provisional Patent: US 63/016,446. Filed 5/5/2020.

\item Charles Schaper, ``Synthesis of Genetic Structures", Provisional Patent: US 63/027,320. Filed  5/19/2020.

\item Charles Schaper, ``Design of DNA, Genetic Codes, and Life Function", Provisional Patent. Ref: E20206R233375951. Filed 6/28/2020.

\end{enumerate}

\clearpage
\newpage

\section{Description of Book Contents} \label{app:book}
The results described in this article are expanded and go into more detail, including new representations and results, especially related to  translation of amino acids into proteins via the double helix in conjunction with steroid molecules as described for the first time in \cite{schaper2020design}.  Chapters comprising this book  (www.molecularprimer.com) include:
\begin{itemize}
   \item Discovering the Primer of DNA - The basic discovery process that I made to identify the common structure of a steroid molecule embedded within each DNA nucleotide pair, the structural match of steroid hormones to DNA structure, and its correlation in function of an intermolecular bond to pharmacological efficacy.
   \item Encoding DNA - The encoding of a unified complex, originally synthesized through intermolecular coupling of a pair of hydrogen bonded steroid structured molecules is described.
   \item Transmitting DNA - The transmission of the encoded complex through the formation of a DNA double helix and steroid molecules which provide access to the information content contained within the double helix.
   \item Decoding DNA - The processes of decoding the double helix structure through the function capability provided by the steroid molecules, including decoding tables of an interaction vessel formed by the steroid molecules.
   \item Translating DNA - The mapping of the nucleotide triplet to amino acid is shown through the analysis of the structural and chemical characteristics of the DNA double helix formed in conjunction with the steroid molecules.
   \item Example - An example is provided of constructing a protein chain of seven amino acids, including the encoding, transmission, decoding and translation aspects.
   \item Replication - Replication of the double helix through the steroid molecules is shown, along with error correction procedures.
\end{itemize}

\end{appendices}

\clearpage
\newpage


\end{document}